\documentclass{aa}
\usepackage{caption}  
\usepackage{graphicx}
\usepackage{amsmath}
\usepackage{amssymb}
\usepackage{subcaption}
\usepackage{lscape}
\usepackage{longtable}
\usepackage{rotating}
\usepackage{diagbox}
\usepackage{upgreek}  
\usepackage{gensymb}    
\usepackage[multiple]{footmisc} 
\usepackage{txfonts}
\usepackage[colorlinks=true,linkcolor=blue,citecolor=blue]{hyperref}
\usepackage{xcolor}

\usepackage{epstopdf}
\usepackage[switch, modulo]{lineno}
\begin{document}

   \title{Far-infrared to centimeter emission of very nearby galaxies with archival data}
   \titlerunning{FIR-cm emission of very nearby galaxies}

   \author{L.Correia \inst{1}, C. Bot\inst{1}, J. Chastenet \inst{2}, A.Rymar\inst{1}, R.Paladini \inst{3}, M. Bethermin\inst{1}, D. Ismail\inst{1}, K.A.Lutz \inst{4}, J.-P. Bernard\inst{5}, A. Hughes\inst{5}, D. Paradis\inst{5}, N. Ysard\inst{5} \fnmsep\
          }
    \authorrunning{L. Correia et al.}

   \institute{Université de Strasbourg, CNRS, Observatoire astronomique de Strasbourg, UMR 7550, F-67000 Strasbourg, France \\
              \email{lucie.correia@astro.unistra.fr}
        \and
             Sterrenkundig Observatorium, Ghent University, Krijgslaan 281-S9, 9000 Gent, Belgium  
        \and
            IPAC, California Institute of Technology, Pasadena, CA., 91125, USA 
        \and 
            DLR Galileo Competence Center, Oberpfaffenhofen, 82234 Wessling, Germany
        \and
        Institut de Recherche en Astrophysique et Planétologie, Toulouse, France
            }

   \date{Received September 15, 1996; accepted March 16, 1997}

 
\abstract
{Compared to the well-studied infrared and radio domains, galaxy emission from a few millimeter (mm) to the centimeter (cm) range (300--30~GHz) has been less observed. In this domain, galaxy emission consists of dust thermal emission, superimposed on free-free and synchrotron emission with a possible additional contribution from anomalous microwave emission (AME) that peaks near 1~cm. 

The aim of this study is to accurately characterize the integrated spectral energy distribution (SED) of galaxies in the mm--cm range.

We used COBE-DIRBE, IRAS, \textit{Planck}, and WMAP all-sky surveys, brought to the same spatial resolution of $\sim 1\degree$, to cover 18 photometric bands from 97 $\mathrm{\upmu}$m to 1.3 cm. Given the low angular resolution and mixing with foreground and background emission that hampers the detection of the galaxy, our sample consists of 6 of the brightest, nearby galaxies: LMC, SMC, M31, M33, NGC~253 and NGC~4945.  We subtract  Milky Way dust emission, distant unresolved galaxies, and foreground point sources in the fields. We fit each integrated SED with a model of thermal dust, free-free, synchrotron, AME and Cosmic Microwave Background (CMB) temperature fluctuations.
The integrated SEDs of our sample of galaxies are well fitted by the model within the uncertainties, although degeneracies between the different components contributing to the mm--cm emission complicate the estimation of their individual contributions. We do not clearly detect AME in any of our target galaxies, and AME emissivity upper limits are weak compared to Galactic standards, suggesting that the signal of AME might be diluted at the scale of a whole galaxy. We infer positive CMB fluctuations in the background of 5 out of our 6 galaxies. This effect might be related to the degeneracy between the dust emissivity index and CMB fluctuations in the background, or linked to the specific spatial distribution of CMB fluctuations coupled with our $1\degree$ resolution and small number statistics.}


\maketitle
\section{Introduction} \label{intro}

The emission of a galaxy is dominated by stellar light in the ultraviolet to near-infrared, by dust emission in the mid- and far-infrared, and by free-free and synchrotron emission in the radio domain. Synchrotron and free-free emission are typically represented as the superposition of two power-law spectra. Dust emission in the far-infrared (FIR) is the thermal emission of large dust grains at thermal equilibrium with the radiation field, and is often represented as a modified black-body spectrum or a sum of modified black bodies. 

Simple dust emission laws have been challenged by observations in the last decades. In particular, observations in the IR to sub-mm have shown breaks or a flattening in the dust emissivity law at long wavelengths in galaxies that has been sometimes described as a sub-mm excess above a given fixed emissivity power-law index dust model \citep[e.g.,][]{lisenfield_2002,bendo_2006,Gordon}. 
Changes in the sub-mm emissivity could be linked to changes in optical properties of dust grains with environmental conditions \citep[e.g.,][]{Demyk_2022}. Based on these laboratory measurements, \citet{Ysard_2024} have demonstrated that the corresponding variations in the intrinsic optical properties of the grains are sufficient to reproduce the range of emissivity spectral index values observed in the diffuse interstellar medium of the Milky Way. Theses changes could also simply be interpreted as revealing the cold dust of galaxies \citep[e.g.,][]{remy-ruyer-2013}. 
A particularly flat dust emissivity was shown to extend to mm wavelengths \citep[e.g.,][]{hermelo2013, tibbs}. This effect was observed in the SMC and LMC up to the cm range \citep{israel_2010,bot2010,planck_2011}, therefore excluding cold dust as a possible origin. Positive Cosmic Microwave Background (CMB) temperature fluctuations in the background can explain the cm excess in the LMC but not fully in the SMC. The emission process to produce such a long-wavelength emission excess above classical dust models is still unclear, but possibilities include magnetic dipole emission from magnetic nanoparticles \citep{draine_2012}, specific emission properties of amorphous solids \citep{meny2007}, or spinning dust emission \citep{bot2010}. 

The fact that CMB fluctuations can be a significant background emission component comparable to the emission of these galaxies is primarily due to the low-surface brightness of SMC and LMC. These galaxies have a lower metallicity than the solar neighborhood reference and are dust poor \citep{Clark_2021,Roman-Duval_2022,Clark_2023}. Furthermore, the thermal dust Rayleigh-Jeans tail is exponentially fainter at long wavelengths, and because of the several degrees angular size of the SMC and LMC on the sky, this emission is spread across large areas. As a consequence, the mm-to-cm ($\sim$ 300-30 GHz) dust brightness of these galaxies is comparable to the emission of CMB fluctuations. Studies of M31 or M33 have also shown the impact of CMB fluctuations on the observed SED \citep{planck2015,tibbs,harper_2023}. While not dominating the mm--cm emission of the galaxy, CMB fluctuations in the background have to be taken into account in order to study the emission of the galaxy itself. An open question is whether dust emission with a flat emissivity index could be erroneously interpreted as part of the background CMB fluctuations. This possibility is supported by the fact that previous mm-to-cm SED studies of the SMC, LMC, and M31 have all required a positive CMB fluctuation in the background of these galaxies. Furthermore, some recent studies have suggested that an unmodeled emission component associated with  galaxies could bias current CMB fluctuations maps \citep{luparello_2022,Hansen_2023,lambas_2024}.

In addition to thermal dust, free-free, and synchrotron emissions in the mm--cm range, a specific component of dust emission is the Anomalous Microwave Emission \citep[AME,][]{dickinson_2018}. The AME peaks around 1~cm (30~GHz), and has been known for several decades \citep[e.g.,][]{Kogut_1996,Leitch_1997}. Extensively observed and studied in our own galaxy \citep[e.g.,][]{watson2005,planck_2014,poidevin2019, cepeda, fernandez-torreiro}, the currently favored emission mechanism is electric dipole radiation from small spinning dust grains in the ISM \citep{Draine_1998_a, Draine_1998_b, Ali_Haimoud_2009,Bell_2019,Casassus_2021,Ysard_2022}.
Extragalactic detections of AME remain rare and, when reported, are confined to localized regions within galaxies, such as a star-forming region in NGC~6946 \citep{murphy2010,hensley2015} and a compact radio source associated with NGC~7425 \citep{murphy2018,murphy_2020}. Additional extragalactic detections are essential for constraining how AME varies across different galactic environments. \citet{Poojon_2024} detected AME in NGC 2903 and marginally in NGC 2146, with higher-frequency, stronger spinning dust emission, consistent with predictions for denser environments like molecular clouds and PDRs. The only detection of AME in the integrated emission of a galaxy so far is in M31 \citep{planck2015,battistelli2019}, although the significance of the detection varies depending on the data used \citep{poidevin2019,harper_2023}. The presence of AME was also studied in the integrated emission of the galaxies M82, NGC~253, and NGC~4945 by \citet{peel}, M33 by \citet{tibbs}, NGC~3627, NGC~4254, NGC~4736, and NGC~5055 by \citep{bianchi_2022}, but only upper limits could be established, finding that it is lower than expected based on the ratio of FIR to AME in the Galaxy.

While the IR to sub-mm emission of galaxies has been extensively studied, their mm-to-cm emission remains poorly constrained. Ground-based observations of nearby galaxies in this wavelength range come with good spatial resolution but with large-scale filtering \citep[e.g.,][]{smith_2021}, and require long observations and precise processing to detect emission beyond the brightest regions of the galaxy given the current instrument/antenna sensitivities \citep[e.g., IMEGIN large program with IRAM 30m led by S. Madden,][]{Katsioli_2023, Ejlali_2025}. Yet, it is unclear how much dust exists in diffuse outer parts of galaxies and it might be significant. For example, \citet{Menard_2010} suggested that half of the dust in galaxies could be in extended halos. In this context, all-sky surveys done with satellite missions like COBE-DIRBE, IRAS, \textit{Planck}, and WMAP hold the potential to detect every sufficiently bright galaxy and capture their integrated IR-to-microwave emission. However, their low angular resolution leads to mixing with foreground and background emission (including Galactic cirrus, extragalactic sources, and CMB fluctuations) that must be carefully disentangled.

This study aims at characterizing the integrated SED of a sample of nearby galaxies in the mm--cm range. For this, we use COBE-DIRBE, IRAS, \textit{Planck}, and WMAP archival all-sky surveys, convolved to the same $\sim 1\degree$ resolution and with cirrus emission and point sources subtracted, to obtain the emission of our target galaxies from 97~$\mathrm{\upmu}$m to 13~mm in 18 photometric bands. We complement these SEDs with archival radio flux densities from the literature and model the resulting SEDs with a simple model of thermal dust emission, free-free, AME, and synchrotron for the galaxy, as well as CMB fluctuation emission in the background.
In principle, any galaxy or point source detected across all bands could be studied, but for nearby galaxies such detections are challenging due to limited resolution and uncertainties in foreground–background subtraction. For this paper, we focus on 6 bright, nearby galaxies and listed in Table \ref{table_gal}). These galaxies were previously studied with \textit{Planck} but the results are difficult to compare because of differences in the methodologies and datasets employed. We excluded M82 from our sample due to strong contamination from the Milky Way foreground at the common $\sim 1\degree$ resolution. M31, M33, LMC, and SMC are part of the Local Group, while NGC~253 and NGC~4945 are more distant. NGC~253 and NGC~4945 are both highly active spiral galaxies, characterized by intense starburst star formation in their central region. NGC~4945 also hosts an obscured Seyfert 2-type active nucleus.

\begin{table}
\centering   
\caption{Celestial coordinates and distances for the sample of galaxies studied in this work.\label{table_gal}}
\begin{tabular}{c|ccc}      
Galaxy & $\alpha$ (J2000) & $\delta$ (J2000) & d [Mpc]\\
\hline
SMC & $00^{h}52^{m}38^{s}$ & $-72^{\circ}48^{'}01^{''}$ & 0.06\\
LMC & $05^{h}23^{m}35^{s}$ & $-69^{\circ}45^{'}22^{''}$ & 0.049\\
M31 & $00^{h}$$42^{m}$$44^{s}$ & $+41^{\circ}16^{'}07^{''}$ & 0.785\\
M33 & $01^{h}$$33^{m}$$51^{s}$ & $+30^{\circ}39^{'}36^{''}$ & 0.809\\
NGC~253 & $00^{h}47^{m}33^{s}$ & $-25^{\circ}17^{'}20^{''}$ & 3.215\\
NGC~4945 & $13^{h}05^{m}27^{s}$ & $-49^{\circ}28^{'}04^{''}$ & 4.61\\
\end{tabular}
\tablefoot{Distances are taken from \citet{Munoz_2018,Richter_1987,McConnachie_2005,Springbob_2005,Allison_2014}.}
\end{table}

The paper is organized as follows. Section \ref{data} presents the data used in our analysis. Section \ref{data_processing} describes the processing to obtain the maps of the galaxies for the study, including the foreground and background subtraction. In Section \ref{emission_models}, we describe our model of dust, free-free, synchrotron, AME emissions for the studied galaxy and CMB fluctuations in the background. Results are presented in Section \ref{results} and discussed in Section \ref{discussion}. Finally, our conclusions are presented in Section \ref{conclusion}.


\section{Data\label{data}}

In this study, we used archival data from all-sky surveys to cover the emission from the IR peak to cm wavelengths, complemented by radio flux densities that we compiled from the literature.

\subsection{IRAS}
The Infrared Astronomical Satellite \citep[IRAS;][]{iras_1984}, launched in 1983, mapped the full sky at 12, 25, 60, and 100~$\mathrm{\upmu}$m. For this study,we used the 100~$\upmu$m IRIS maps from \citet{IRIS} to benefit from improved zodiacal light subtraction, calibration, and zero-level accuracy, allowing us to model the peak of thermal dust emission. We adopted a calibration uncertainty of 13.5$\%$ and a beam width of $\approx$ 4' \citep{Hauser_1998,IRIS}.

\subsection{COBE-DIRBE}
The Cosmic Background Explorer \citep[COBE; ][]{cobe}, launched in 1989, mapped the sky with three instruments from 1.2 to 240~$\upmu$m. We used 3 bands of DIRBE\footnote{\label{fn:site}\url{https://lambda.gsfc.nasa.gov/}} at 100, 140, and 240~$\mathrm{\upmu}$m to further constrain the peak of the thermal dust spectrum, and with zodiacal light already subtracted. The uncertainties in the gain calibration for these bands are 13.5, 10.6, and 11.6 $\%$ for the bands at 100, 140, and 240~$\mathrm{\upmu}$m, respectively \citep{Hauser_1998}. Although the DIRBE beam is significantly non-Gaussian, as shown by \citet{Cambresy_2001}, due to multiple scan directions, the data products we use can be approximated by a rotation-averaged Gaussian beam with a width of $1^{\circ}$.

\subsection{WMAP}
The Wilkinson Microwave Anisotropy Probe (WMAP), launched in 2001, mapped the sky in 5 bands: 3.2, 4.9, 7.3, 9.1, and 13~mm (93.5, 60.7, 40.7, 33.0, and 22.8~GHZ respectively) at a resolution between 12.6' for the lower wavelength and 55.8' for the higher wavelength. We used final nine-years data delivery maps\footref{fn:site} \citep[DR5;][]{Bennett_2013}.  As the instrumental noise is not a limiting factor, we assume a $3\%$ overall calibration uncertainty, as used in other studies \citep{planck_2014}. 

\subsection{Planck}\label{planck_cmb_maps}
The \textit{Planck} satellite, launched in 2009, observed the sky with two instruments, the High Frequency Instrument (HFI; \citet{hfi_2013}), covering the bands at 350, 500, 850~$\mathrm{\upmu}$m, and 1.3, 2.1, 3~mm (857, 545, 353, 217, 143, 100~ GHz, respectively) and the Low Frequency Instrument (LFI; \citet{lfi_2013}) covering three bands centered at 4.3, 6.8, and 10~mm (70, 44, and 30~GHz respectively). The angular resolution is between 5' at the highest frequency and 30' for the lowest \citep{planck_satellite} We adopt calibration uncertainties of 7$\%$ for HFI bands \citep{hfi_2013} and 3$\%$ for LFI bands \citep{lfi_2013}. We used data from the last release "\textit{Planck} Public Release 4" published in 2020. 

We also used some of the \textit{Planck} data products obtained from component separations and downloaded from \textit{Planck} Legacy Archive\footnote{\url{https://pla.esac.esa.int/}}: COMMANDER CMB fluctuation maps \citep{planck_diffuse_2020}, CIB \citep{cib_map} and CO maps \citep{planckco}. These maps were used to estimate foreground dust emission associated with molecular clouds, background emission from distant unresolved galaxies or to remove a first CMB fluctuation estimate around our galaxies to estimate the background zero level emission (see Section \ref{bkg_subtraction}).

\subsection{Radio data}\label{sec:radio}

For each galaxy studied, we gathered radio flux densities from the literature by using the NASA/IPAC Extragalactic Database\footnote{\url{https://ned.ipac.caltech.edu/}} (NED) and VizieR SED photometry services\footnote{\url{https://vizier.cds.unistra.fr/}}. 
We prioritized flux densities obtained with single-dish radio telescopes over those from interferometric observations. This choice was motivated by the idea that single-dish measurements are more sensitive to diffuse and extended emissions while interferometric data might resolve them out. We focused exclusively on data in the radio domain down to 0.3 GHz, which reflects both the limited availability of high-resolution data at lower frequencies and the known turnover of the synchrotron spectrum below $\sim$ 1 GHz, where a simple power-law approximation becomes unreliable. Restricting the analysis to this frequency range ensures robust constraints on synchrotron and free-free emission models. The radio data used in this study are listed in the Appendix for each galaxy (from Table \ref{table_radio_data_lmc} to Table \ref{table_radio_data_ngc4945}).

\subsection{Literature IR-mm data}

For consistency and comparison, we included IR–mm data from the literature, including both similar resolution and higher-resolution observations (e.g. from \textit{Herschel}, IRAM 30 m). These data are not used in our modeling, but we use them for comparison to our fiducial measurements, and to check whether there is evidence for extended emission in the integrated SEDs. 

\subsection{Ancillary data}

In order to assess the contribution of foreground emission from dust in the Milky Way, we use gas maps as templates. 
For the atomic hydrogen gas, we used the HI 4-PI Survey (HI4PI) \citep{HI4PI}, built from the Effelsberg-Bonn HI Survey at a resolution of 16.2'. Because we are interested in using these maps for the Milky Way foreground only, we take the integrated HI column density maps corresponding to velocities between $-$90 and 90 km/s. 

To trace the ionized gas, we used the \citet{finkbeiner} full-sky H$\alpha$ map, which combines reprocessed Virginia Tech Spectral line Survey (VTSS), Southern H-Alpha Sky Survey Atlas (SHASSA), and Wisconsin H-Alpha Mapper (WHAM) Northern Sky Survey data. VTSS and SHASSA were calibrated to the WHAM zero point and reprocessed to remove artifacts, producing a well-sampled map at 6' FWHM.


\section{Data processing}\label{data_processing}

\subsection{Combination of all-sky surveys }\label{combination_archival_data}

The combination of IRAS, DIRBE, \textit{Planck}, and WMAP maps allows us to explore nearby galaxy emission in 18 photometric bands, from the peak of thermal dust emission in the FIR up to cm wavelength, a wide frequency coverage.

These data are encoded in the HEALPix format (Hierarchical Equal Area isoLatitude Pixelisation; \citealt{Healpix}). In order to be directly comparable pixel-to-pixel, all maps were smoothed to a common resolution of $1^{\circ}$ using the smooth function of the healpy package, applying a Gaussian approximation. This resolution corresponds to the lowest of our sample, defined by the 13 mm band of WMAP and the DIRBE data. Given the low angular resolution, the sampling was reduced to a consistent HEALPix grid of $N_{\rm side}$ = 256 (pixel angular size of $\sim$ 13.7'). 

All maps were converted to the same unit, MJy/sr, using conversion factors from $K_{\rm cmb}$ to MJy/sr listed \citep{lagache2020} for \textit{Planck}  and in the \textit{Planck} Sky Model for WMAP\footnote{\textit{Planck} Sky Model User Manual: \url{https://apc.u-paris.fr/~delabrou/PSM/Public/PSM_user_manual_v1_7_8.pdf}}. 

\subsection{Foreground and background emission subtraction}\label{bkg_subtraction}

At the low angular resolution of this study ($1\degree$), foreground and background emission superimpose and mix together with the emission of the galaxy we want to study: zodiacal light, Milky Way dust emission, other point sources close in projection to the galaxy, unresolved galaxies in the background and CMB fluctuations all emit at IR to radio wavelengths. While some of these emission components like zodiacal light are sufficiently large scale or faint to be ignored, the other ones can be significant or even brighter than the emission of the galaxy itself. In this study, we attempt to remove these emissions using template maps, except for CMB fluctuations that we model together with the studied galaxies. Here we briefly describe the overall process for foreground and background subtraction.

\subsubsection{Cosmological Microwave Background (CMB) fluctuations}

At mm-to-cm wavelengths, CMB fluctuations contribute significantly and may even dominate the total observed signal in the mm--cm regime. Different CMB temperature fluctuation maps based on different component separation methods exist \citep{cmb_maps,planck2020}. Previous analyses of nearby galaxies \citep[e.g.,][]{tibbs,harper_2023} have shown that the choice of the component separation map that is used for CMB temperature fluctuations can change the resulting physical parameters estimated for the galaxy. Furthermore, these CMB fluctuation maps cannot always be trusted in the direction of our studied galaxies and are imprinted or masked in the maps that are openly available. This is due to the fact that CMB fluctuations signal can be partially degenerate with dust and radio emission, complicating their removal \citep[e.g.,][]{planck_2011,bobin2016}. In this paper, we model CMB fluctuations jointly with the galaxy emission components. We use the COMMANDER CMB maps \citep{planck2020} for visualizing the CMB fluctuations,  and for estimating a CMB-free background level around each galaxy, but CMB fluctuations are not removed from the integrated SEDs that we analyze. We adopt the COMMANDER PR4 products for consistency with the PR4 intensity maps and because they perform best in regions strongly contaminated by Galactic dust and other foregrounds \citep{planck2020}.

\subsubsection{Foreground and background source removal}
Other galaxies and compact sources in our fields can bias our background estimates, foreground subtraction, and create additional scatter in the dust–gas correlation that we use to remove extended foreground dust emission. To address this, we subtract all resolved sources located within the fields of our targets using the source catalogs of each all-sky survey: the COBE/DIRBE Point Source Catalog \citep{cobe_dirbe_source_catalogue}, the IRAS Galaxies+QSOs Catalog \citep{iras_catalogue}, the \textit{Planck} PCCS2 and PCCS2E \citep{pccs2}, and the WMAP Nine-Year Source Catalog \citep{wmap_source_catalogue}.

In the case of M31, a source located close to the galaxy affects the measured fluxes and, consequently, the shape of its integrated SED. The analysis by \citet{harper_2023} is more accurate in this respect, as they rely on higher-resolution studies to correct for this effect. In our work, we aimed to apply a consistent methodology across all galaxies, and we do not have comparable high-resolution data for the entire sample. Therefore, we adopted a uniform approach for all galaxies, which does not specifically correct for this source.

For each photometric band, we generate a source map on a Healpix grid at the survey resolution by projecting catalog fluxes at the listed positions and converting them to MJy/sr. Catalogs often provide multiple flux estimates depending on the extraction method. For \textit{Planck}, GAUFLUX is adopted for extended sources and DETFLUX for compact ones. We consider sources as extended if the geometric mean of the fitted ellipse exceeds 1.5 times the beam FWHM \citep{pccs2}. We take care not to subtract emission that is intrinsic to our target galaxies. Specifically, we exclude all catalog sources within $2^{\circ}$ of the SMC and M31, $5^{\circ}$ of the LMC, and $0.1^{\circ}$ of M33, NGC~253, and NGC~4945. This ensures that the entire emission of these galaxies, which may be divided into several sub-sources in some catalogs, is not inadvertently subtracted.

The resulting source maps are smoothed to $1^\circ$, degraded to $N_{\rm side}=256$, and subtracted from the intensity map for each photometric band.

\subsubsection{Zero level estimation and CIB removal}\label{sec:CIB}

The Cosmic Infrared Background (CIB) arises from the integrated emission of unresolved extragalactic sources, including distant galaxies and quasars. Although it contributes to the overall FIR intensity maps, its effect is minimized in our analysis through local background subtraction around each region of interest. Consequently, the absolute zero level of the maps does not influence our results.

In addition to a constant background, CIB emission has also anisotropies, reflecting both the large-scale clustering of the contributing sources and their intrinsic luminosity variations \citep{planck_cib}. To estimate the spatial distribution of the CIB, we inspected the CIB Planck maps at 350, 500, and 850~$\upmu$m in MJy/sr \citep{planck_cib}, but we do not find these fluctuations to be significant for our study.

\subsubsection{CO line contribution to photometric bands}

\textit{Planck} maps at 1.3 and 3 mm (217 and 100 GHz) can contain significant J = 2 $\rightarrow$ 1 or J = 1 $\rightarrow$ 0 CO line emission \citep{planckco} on top of the dust emission, from foreground Milky Way interstellar gas or from the molecular gas of the galaxy we study. Although these contributions can be estimated from \textit{Planck} CO maps \citep{planckco},  we chose to omit the 1.3 and 3 mm bands from the SED fitting to avoid bias.

\subsubsection{Milky Way dust emission foreground}

Dust in the Milky Way is emitting in every region of the sky, even at high Galactic latitude where our galaxies are located. We therefore need to remove this dust emission in the foreground of the galaxies we want to study. The foreground emission is more prominent in the FIR but extends to all wavelength bands \citep{planckxvii_2014}. Because gas and dust are correlated, we can use gas tracers to estimate the dust emission foreground and subtract it to the all-sky surveys we use. In section \ref{sec:CIB}, we  already used this correlation at high Galactic latitude to define a first order background emission for the sky. Here, we want to use gas maps as templates for the dust emission in the foreground. 
The idea is to estimate dust-gas emissivity factors, apply these to the gas map to compute a template of foreground dust emission in each band and remove it in front of our studied galaxies.

Although \citet{planckxvii_2014} estimated well-constrained dust-to-HI ratios for the diffuse and high-latitude sky, variations of these ratios around their average values exist on the sky \citep{Bot_2009,planck_2011diffuse,planckxvii_2014} and correspond to actual variations of dust properties \citep{Ysard_2024}. In order to take into account these variations in the dust emissivity and temperatures of the foreground, we chose to perform a local estimate of the cirrus foreground around each galaxy. Further details are provided in Appendix \ref{appendix_dust_emission}.

\begin{figure*}[!t]
    \centering
    \begin{subfigure}[t]{0.49\textwidth}
        \centering
        \includegraphics[height=0.8\textheight]{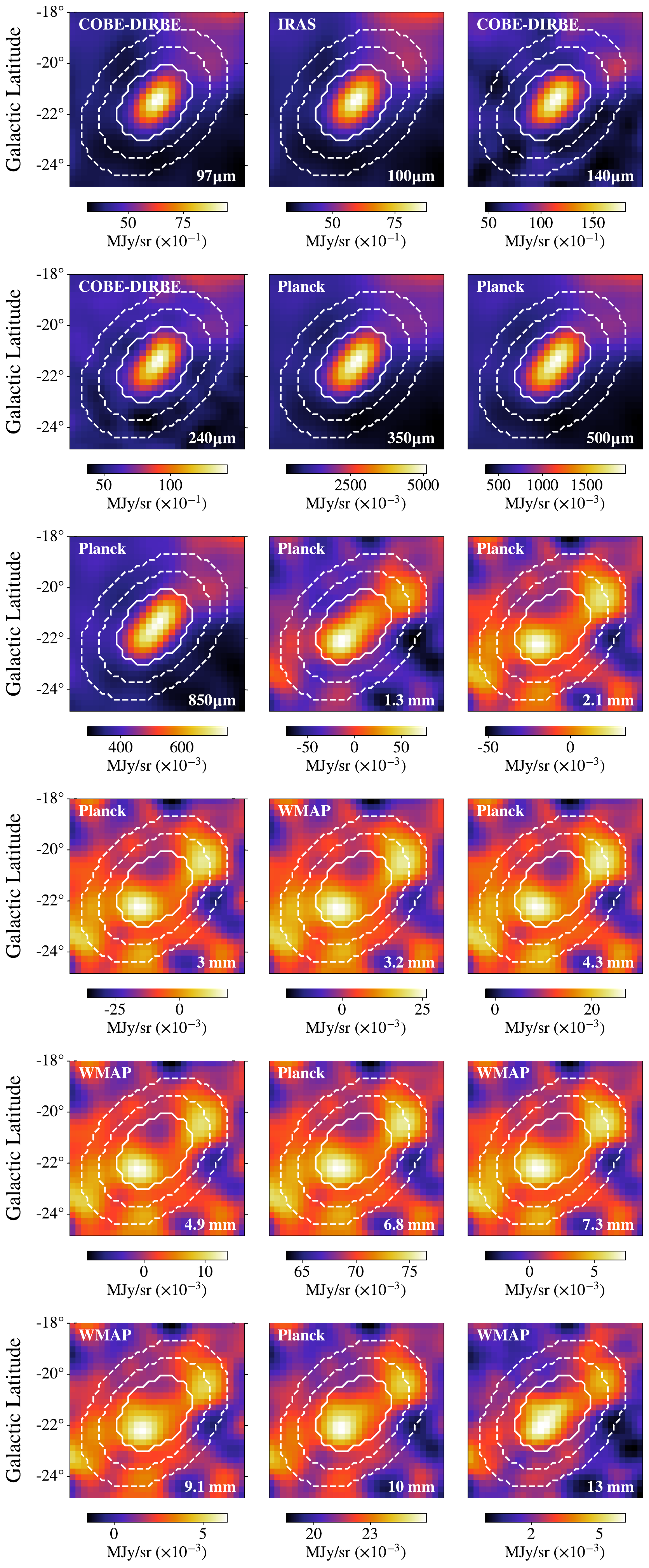}
        \caption{}
        \label{map_m31_with_cmb}
    \end{subfigure}
    \hfill
    \begin{subfigure}[t]{0.49\textwidth}
        \centering
        \includegraphics[height=0.8\textheight]{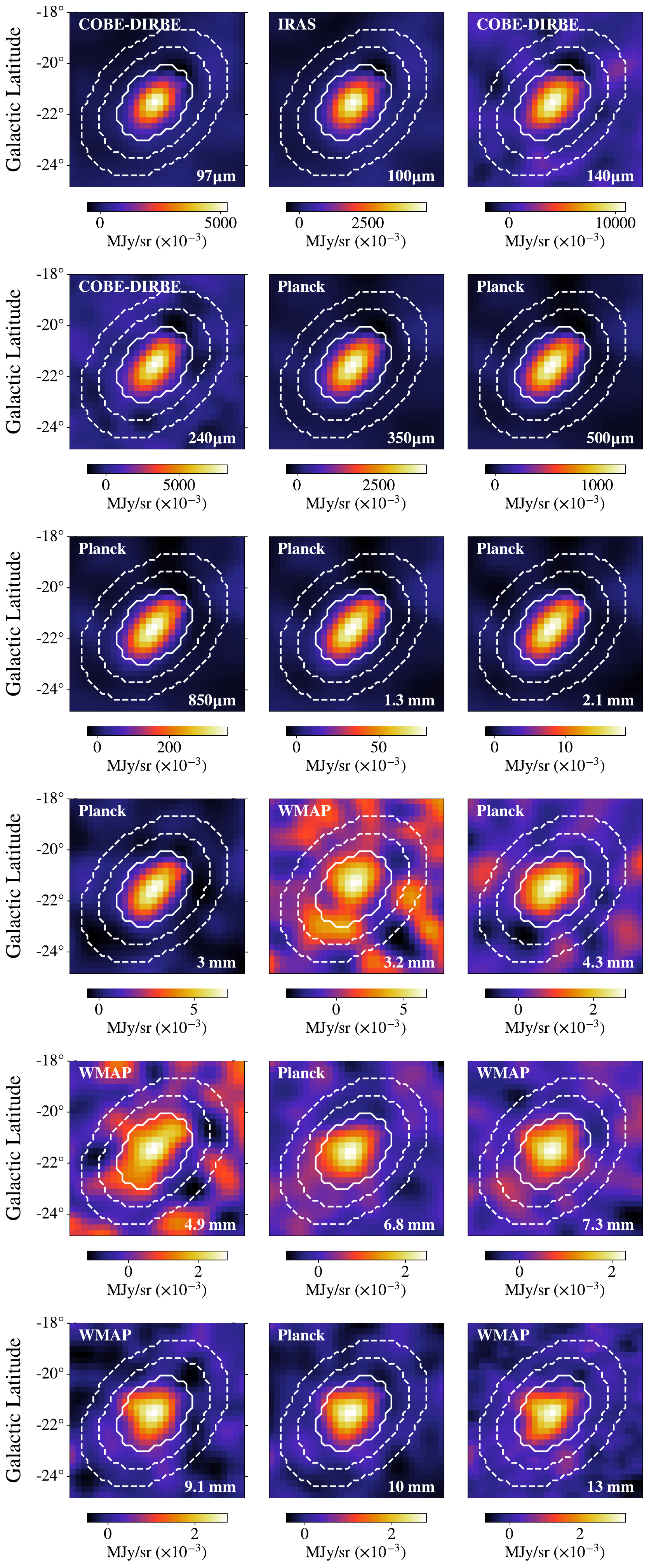}
        \caption{}
        \label{mapm31nocmb}
    \end{subfigure}
    
    \caption{(a) Maps of M31 at $1^{\circ}$ of resolution, in 18 photometric bands of COBE-DIRBE, IRAS, \textit{Planck} and WMAP from 97 $\mathrm{\upmu}$m to 13 cm, \textit{before} any subtraction of foreground and background emission. (b) Maps around M31 \textit{after} the subtraction of foreground and background sources. The solid and dashed white lines delineate the on-source and background regions used for the aperture photometry, respectively.}
    \label{map_m31_sidebyside}
\end{figure*}

\subsubsection{Resulting foreground/background subtracted maps}

Figures \ref{map_m31_with_cmb} and \ref{mapm31nocmb} show the maps of M31 as an example before and after subtraction of the CMB fluctuations, CIB, individual sources and dust in the foreground as presented in this section. Comparing with Figure \ref{map_m31_with_cmb}, we can see that the galaxy in Figure \ref{mapm31nocmb} is therefore well detected, even at long wavelengths. At 3.2 and 4.9~mm, the data are significantly noisier, limiting the signal quality in these bands. The maps after the subtraction of the foreground and background emissions for the other galaxies of the sample are presented on the public platform \href{https://doi.org/10.5281/zenodo.17662333}{Zenodo}.

\subsection{Projection and aperture photometry\label{sec:projandaper}}

After subtraction of the different foreground and background emission, for each galaxy of our sample, we project all maps on a $30\times 30$ pixel grid centered on each galaxy and with a pixel size of 13.74'. The LMC and SMC being extended galaxies (respectively $8^{\circ}$ and $2^{\circ}$), we built pixel grids of 100 $\times$ 100 pixels for LMC and 50 $\times$ 50 pixels for SMC.  The galaxy coordinates (and hence central pixels) were taken from the SIMBAD database\footnote{\url{https://simbad.cds.unistra.fr/simbad/}} and are summarized in Table \ref{table_gal}. The maps we obtain and use for each galaxy are presented before and after subtraction for all 18 bands in Figures \ref{map_m31_with_cmb} and \ref{mapm31nocmb} for M31 as an example, and on \href{https://doi.org/10.5281/zenodo.17662333}{Zenodo} for the other galaxies of our sample.

To obtain the integrated flux density of each galaxy, we performed aperture photometry on the galaxy maps in all bands. The emission at the galaxy's position is determined by measuring the flux density within a region centered on the galaxy ("ON" region) and subtracting the background emission estimated in an annulus surrounding the galaxy ("OFF" region). For the SMC, LMC and M31, we chose the same apertures as in \citet{planck_2014} and \citet{planck2015}. For M33, NGC~253 and NGC~4945, we take a circular region  with a radius of 80' surrounded by an annulus of radius 42' that is offset further out. The ON and OFF regions we use are delineated by the solid and dashed white lines, respectively.

We emphasize that we have chosen to measure the integrated flux density of the galaxy in the central region, on the convolved and foreground/background subtracted maps, except for the CMB fluctuations that are kept in and will be taken into account in the modeling. For the background (OFF) region, we take the median of the surface brightnesses on the convolved maps with all foreground and background emissions removed (including the CMB fluctuations) to ensure that the average background level is not mainly driven by the CMB temperature fluctuations.

Most of the uncertainty on the integrated fluxes of the galaxies comes from remaining variations due to the imperfect foreground and background emission subtractions. We quantify this source of uncertainty by computing the standard deviation of the surface brightnesses observed in the background region around each galaxy, as measured in maps with all foreground and background emission subtracted (including the CMB fluctuations). We then add these uncertainties quadratically to the calibration gain uncertainties from the instruments.

The resulting flux densities in all 18 bands from $\sim 100~\upmu$m to 1.3~cm for each galaxy are listed in Table \ref{table_flux_cmb} and plotted as a function of frequency with gray points in Figures \ref{sed_m31}, \ref{seds_lmc_smc}, \ref{seds_m33}, and \ref{seds_ngc253_ngc4945}. These obtained integrated SEDs are complemented with the radio data from the literature as described in Section \ref{sec:radio} and listed in appendix \ref{sec:radiodata}. 
These flux densities constitute a complete set of integrated SEDs from the IR to the radio domain, with excellent mm--cm coverage. They include both the galaxy emission and background CMB fluctuations, which we modeled jointly and analyzed in the following sections.

\begin{table*}
\caption{Integrated flux densities (in Jy) obtained for each galaxy from the foreground and background subtracted maps but keeping CMB fluctuations in.\label{table_flux_cmb}}            
\centering 
\begin{tabular}{|c|c|c|cccccc|}     
\hline                       
Mission & $\lambda$($\mathrm{\upmu}$m) & $\nu$(GHz) & LMC & SMC & M31 & M33 & NGC~253 & NGC~4945 \\
\hline
COBE & 97   & 3000 & $156000 \pm 44000$ & $13400 \pm 3800$ & $2290 \pm 652$ & $972 \pm 276$ & $2050 \pm 583$ & $1783 \pm 506$ \\
IRAS & 100  & 3000 & $182000 \pm 25600$ & $14000 \pm 2420$ & $2730 \pm 474$ & $990 \pm 284$ & $1690 \pm 270$ & $1770 \pm 946$ \\
COBE & 140  & 2141 & $190000 \pm 51500$ & $13100 \pm 3550$ & $5006 \pm 1360$ & $1287 \pm 349$ & $2190 \pm 595$ & $2108 \pm 572$ \\
COBE & 240  & 1250 & $112000 \pm 31000$ & $8390 \pm 2310$ & $3927 \pm 1080$ & $1260 \pm 347$ & $969 \pm 267$ & $1520 \pm 419$ \\
\textit{Planck} & 350  & 857  & $71900 \pm 8470$ & $6300 \pm 1360$ & $2679 \pm 237$ & $636 \pm 146$ & $481 \pm 65$ & $850 \pm 618$ \\
\textit{Planck} & 500  & 545  & $24600 \pm 3100$ & $2570 \pm 463$ & $920 \pm 85$ & $239 \pm 51$ & $132 \pm 21$ & $248 \pm 210$ \\
\textit{Planck} & 850  & 353  & $7554 \pm 842$ & $899 \pm 130$ & $281 \pm 20$ & $80 \pm 14$ & $35 \pm 5$ & $49 \pm 61$ \\
\textit{Planck} & 1300 & 217  & $2208 \pm 189$ & $266 \pm 30$ & $83 \pm 5$ & $25 \pm 4$ & $11 \pm 1$ & $-12 \pm 14$ \\
\textit{Planck} & 2100 & 143  & $848 \pm 51$ & $99 \pm 7$ & $29 \pm 1$ & $9.6 \pm 0.9$ & $5.2 \pm 0.4$ & $-18 \pm 4$ \\
\textit{Planck} & 3000 & 100  & $479 \pm 21$ & $52 \pm 3$ & $16 \pm 1$ & $5.2 \pm 0.5$ & $3.3 \pm 0.4$ & $-12 \pm 1$ \\
WMAP & 3200 & 93.5 & $417 \pm 24$ & $43 \pm 6$ & $13 \pm 2$ & $3.7 \pm 1.7$ & $3.6 \pm 2.4$ & $-13 \pm 2$ \\
\textit{Planck} & 4300 & 70   & $283 \pm 11$ & $29 \pm 2$ & $7.5 \pm 0.6$ & $2.4 \pm 0.5$ & $1.4 \pm 0.4$ & $-5.7 \pm 0.7$ \\
WMAP & 4900 & 60.7 & $244 \pm 12$ & $23 \pm 3$ & $6.8 \pm 0.8$ & $3.0 \pm 0.9$ & $1.8 \pm 0.8$ & $-3.9 \pm 0.7$ \\
\textit{Planck} & 6800 & 44   & $191 \pm 9$ & $18 \pm 2$ & $3.9 \pm 0.4$ & $1.1 \pm 0.4$ & $1.4 \pm 0.2$ & $-1.8 \pm 0.7$ \\
WMAP & 7300 & 40.7 & $189 \pm 8$ & $17 \pm 2$ & $3.6 \pm 0.4$ & $1.2 \pm 0.5$ & $1.2 \pm 0.5$ & $-1.3 \pm 0.9$ \\
WMAP & 9100 & 33   & $176 \pm 8$ & $15 \pm 2$ & $3.3 \pm 0.3$ & $1.0 \pm 0.4$ & $1.2 \pm 0.4$ & $-0.1 \pm 0.9$ \\
\textit{Planck} &10000& 30   & $169 \pm 9$ & $15 \pm 2$ & $2.9 \pm 0.3$ & $0.99 \pm 0.28$ & $1.3 \pm 0.3$ & $0.13 \pm 0.96$ \\
WMAP & 13000& 22.8 & $173 \pm 9$ & $14 \pm 2$ & $2.7 \pm 0.4$ & $0.82 \pm 0.42$ & $1.4 \pm 0.4$ & $0.69 \pm 1.21$ \\
\hline                
\end{tabular}
\end{table*}

\section{Emission model}\label{emission_models}

Since we expect the galaxy emission in the mm-to-cm range to be a combination of dust, free-free, synchrotron and AME, with the presence of CMB fluctuations in the background, and each of these components have a specific spectral shape, we use a model that is a combination of these different contributions, following the approach of \citet{harper_2023}.

The integrated emission of the galaxy is modeled as:
\begin{equation}\label{eq:model_tot}
\begin{aligned}
    \text{S}_{\text{tot}}(\nu) &= \, \text{S}_{\text{dust}} + \text{S}_{\text{AME}} +\text{S}_{\text{ff}} + \text{S}_{\text{syn}}
      + \text{S}_{\text{CMB}}
\end{aligned}
\end{equation}
where S($\nu$) are flux densities in Jy at a given frequency band. 

\subsection{Dust emission}
Dust emission dominates the FIR–mm regime of galaxies and can be modeled in various ways: from full dust models \citep[e.g.,][]{Draine_2001,jones2013,kohler2014,kohler2015,ysard2015,Siebenmorgen_2023,Hensley_2023,Ysard_2024} and mixing recipes accounting for environmental and heating variations \citep[e.g.,][]{Dale_2001,Dale_2002,Draine_2007,Galliano_2018_herbie}, to more phenomenological approaches \citep[e.g.,][]{Gordon,Chiang_2018}. Here we adopt a simple modified blackbody since (1) the FIR–mm emission is dominated by large grains in thermal equilibrium, (2) we aim to limit the number of free parameters in our multi-component model, (3) the integrated SEDs of nearby galaxies are well described at long wavelengths by this model \citep[e.g.,][]{harper_2023}, and (4) it facilitates comparison with other studies, including at higher redshift \citep[e.g.,][]{ismail_2023}.

Our simple dust emission model is hence described as a single modified blackbody as:
\begin{equation}
    \mathrm{S_{dust}} = \frac{2\,h\,\nu^{3}}{c^{2}}\left(\frac{\nu}{353 ~\mathrm{GHz}}\right)^\beta \tau_{353} \left(e^{h\nu / k_{b}T_{\rm d}}-1\right)^{-1} \Omega
\end{equation}
with the free parameters $\beta$ the dust emissivity index, $\tau_{353}$ the optical depth at 353~GHz, and $T_{d}$ the dust temperature  (in K). $\nu$ is the observed frequency in GHz and $\Omega$ the solid angle of the aperture (in sr).

\subsection{Anomalous Microwave Emission}

The spinning dust component is modeled as:
\begin{equation}
     \mathrm{S_{AME}} (\nu) = C_{\mathrm{AME}} \, S_{\mathrm{spin}}(\nu) \, \Omega
\end{equation}
where we take $S_{\rm spin}$ as the AME emission spectrum in MJy/sr for the diffuse Milky Way dust reference, and the free parameter $C_{\rm AME}$ acts as the scaling factor of AME. As stated in \citet{Ysard_2022}, spinning dust models using carbonaceous nano-grains give comparable results regardless of their exact nature, PAHs as in SpDust or amorphous hydrocarbons as in THEMIS \citep[The Heterogeneous dust Evolution Model for Interstellar Solids,][]{jones2013,kohler2014,jones2014,kohler2015,ysard2015}\footnote{\url{https://www.ias.u-psud.fr/themis/}}, the model that we use. We model the spinning dust component as a fixed template representing the warm neutral medium as defined by \citet{Draine_1998_b} and \citet{Weingartner_2001}, with a fixed peak frequency at 31 GHz.

\subsection{Free-free emission}

Free-free emission, arising from electrons accelerated in the electric fields of diffuse gas and ionized regions \citep{Condon_1992}, exhibits a nearly flat spectrum: at low frequencies its intensity rises roughly as $\nu^{2}$ with an almost constant brightness temperature, whereas at high frequencies the intensity decreases slowly as $\nu^{-0.1}$. Consequently, significant degeneracies can arise between the free-free component and other emissions, particularly synchrotron and AME. We modeled free-free emission as:
\begin{equation}
     \mathrm{S_{ff}} = \frac{2\,k_{b}\,(T_{e}\,(1-e^{-\tau_{ff}}))\,\Omega\,\nu^{2}}{c^2}
\end{equation}
where $T_{\rm ff}= T_{\rm e}\,(1-e^{-\tau_{\rm ff}})$ is the free-free brightness temperature, $T_{\rm e}$ is the electron temperature\footnote{We assume a fixed electron temperature of 8000~K \citep{planck_2014} and singly ionized scattering ions.} and $\tau_{\rm ff}$ the free-free optical depth. The optical depth is defined by \citet{Draine2011}:
\begin{equation}
     \mathrm{\tau_{ff}} = 5.468 \times 10^{-2}\,T_{e}^{-1.5}\,\nu^{-2}_{\mathrm{GHz}}\,EM\,g_{ff} 
\end{equation}
with the emission measure EM in pc \text{cm$^{-6}$}, and the Gaunt factor $g_{\rm ff}$ derived from \citet{Draine2011}. In this free-free emission model, the only free parameter is EM.

\subsection{Synchrotron emission}
Synchrotron emission results from the acceleration of ultra-relativistic electrons within magnetic fields, and its spectral distribution is generally well represented by a power-law function \citep{Condon_1992}. Accordingly, we model the synchrotron emission with a simple power-law:
\begin{equation}
    \mathrm{S_{\text{syn}}} = C_{\text{syn}} \,\nu^{\alpha_{\text{syn}}}
\end{equation}
with two free parameters: ${C_{\rm syn}}$ is the synchrotron amplitude at 1 GHz, and $\upalpha_{\rm syn}$ is the flux density spectral index. 

\subsection{Background CMB fluctuation emission}
We take into account the CMB fluctuations in our model. Their flux densities are modeled as:
\begin{equation}
    \mathrm{S_{CMB}}(\nu) = \frac{2\,h\,\nu^{4}}{c^{3}\,T_{\rm CMB}} \frac{x\,e^{x}}{e^{x}-1} \,\delta_{\text{CMB}} \, \Omega
\end{equation}
with $x=h\nu/k_{b}T_{\rm CMB}$, $\mathrm{T_{CMB}}$ the temperature of the CMB (2.72548 K; \citealt{Fixsen_2009}), and the free parameter $\mathrm{\delta_{CMB}}$ corresponding to the amplitude of CMB temperature fluctuations in K.
Modeling the CMB fluctuations together with the galaxy emission allows us to explore possible degeneracies between free parameters. 

\subsection{Fitting with Markov Chain Monte Carlo}

The SEDs we fit correspond to the emission integrated in a set of 18 photometric bands from the different instruments. For a fair comparison, we integrate our total model (equation \ref{eq:model_tot}) in the same photometric bands as well. We do this using the band-pass filters from each instrument and integration libraries from the DustEM IDL tool\footnote{\url{http://dustemwrap.irap.omp.eu/}} \citep{dustem}.

To fit our emission model (with eight free parameters; $\beta$, $\tau_{353}$, $T_{d}$, $C_{\rm AME}$, $C_{\rm ff}$ , ${C_{\rm syn}}$, $\upalpha_{\rm syn}$ and $\mathrm{\delta_{CMB}}$) integrated into photometric bands, to the observed SED of each galaxy, we use a combination of two fitting methods:

We begin with a PYTHON NUMPY Levenberg-Marquardt \citep{Levenberg_1944,Marquardt_1963} non-linear least squares method to minimize the difference between our model and the observed SED. We provide initial guesses for the model parameters ($\beta \in [0,3]$, $\log_{10}\tau_{353} \in [-8,8]$, $T_{\rm d} \in [10,50]~\mathrm{K}$, $C_{\rm ff} \in [0,10]$,  $C_{\rm syn} \in [0,100]$, $\upalpha_{\rm syn} \in [-5,5]$, $\delta_{\rm CMB} \in [-50,50]~\upmu\mathrm{K}$, and $C_{\rm AME} \in [0,5]$), and we assume additive Gaussian noise with independent errors for each photometric band, using the reported observational uncertainties to weight the fit. This method provides a first estimate of the best-fit parameters and and their associated uncertainties, obtained as the square root of the diagonal elements.

We then used the \texttt{emcee} \citep{emcee_Foreman-Mackey2019} implementation to do a Markov Chain Monte Carlo (MCMC) fitting. This second step allows us to better explore the parameter space and study the potential degeneracies between model parameters. The MCMC process is initialized with the start parameters found by the least-squares fitting. Uniform priors were applied to all parameters, with positivity enforced for all amplitude parameters except the CMB anisotropy. For this work we used 200 chains, that run for 10000 iterations, with a burn-in of 4000 samples to allow the chains to converge. Finally, we thin the chains by taking every 15th sample to remove sample-to-sample correlations, ensuring the final samples are independent for a robust parameter estimation.

While the Levenberg-Marquardt step provides a quick and approximate best-fit solution, the MCMC step gives the full posterior probability distribution for each parameter. The best-fit parameters are reported in Appendix \ref{appendix_best_fit_parameters} and are defined as the median of these posterior distributions. The quoted uncertainties correspond to the 16th and 84th percentiles. The corner-plots, such as the one shown in Figure \ref{cornerplot_m31} for M31, present the posterior probability density estimated by the MCMC. In the diagonal, the panels show the probability distributions of each parameter, while other panels exhibit the joint distributions between parameters, highlighting some degeneracies.

\section{Results}\label{results}

\begin{figure*}
    \centering
    \includegraphics[width=0.5\textwidth]{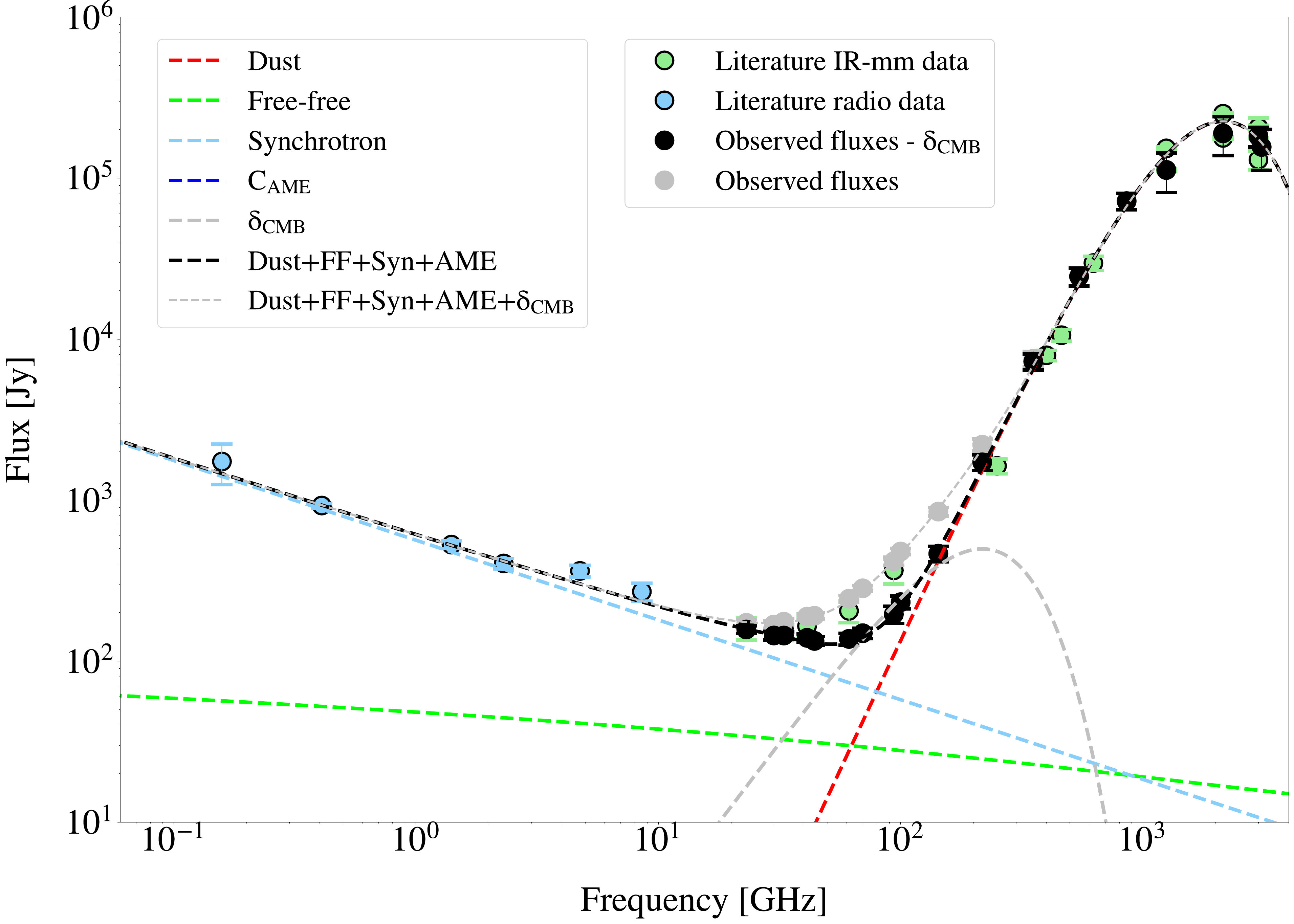}\hfill\includegraphics[width=0.5\textwidth]{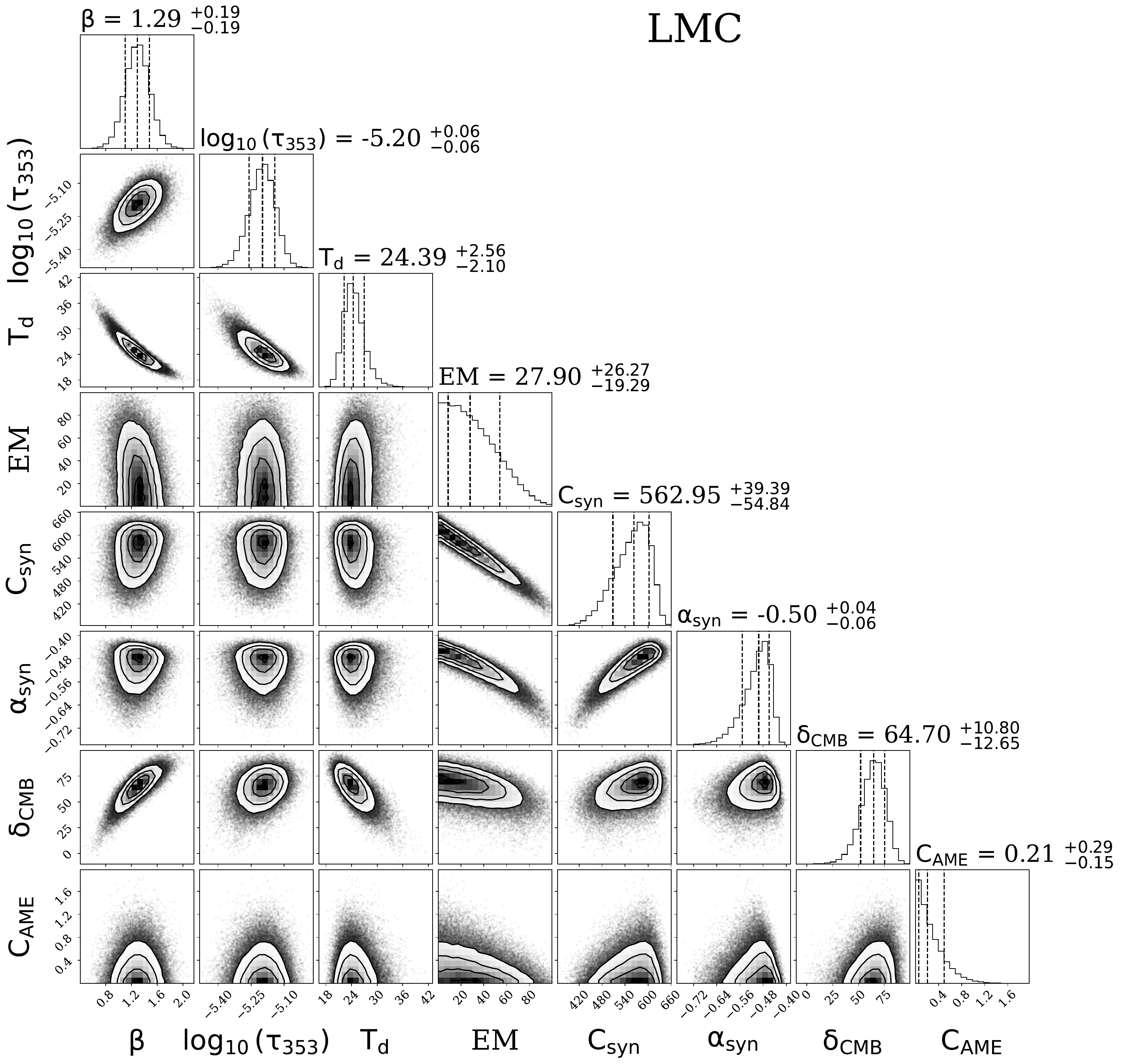}
    \includegraphics[width=0.5\textwidth]{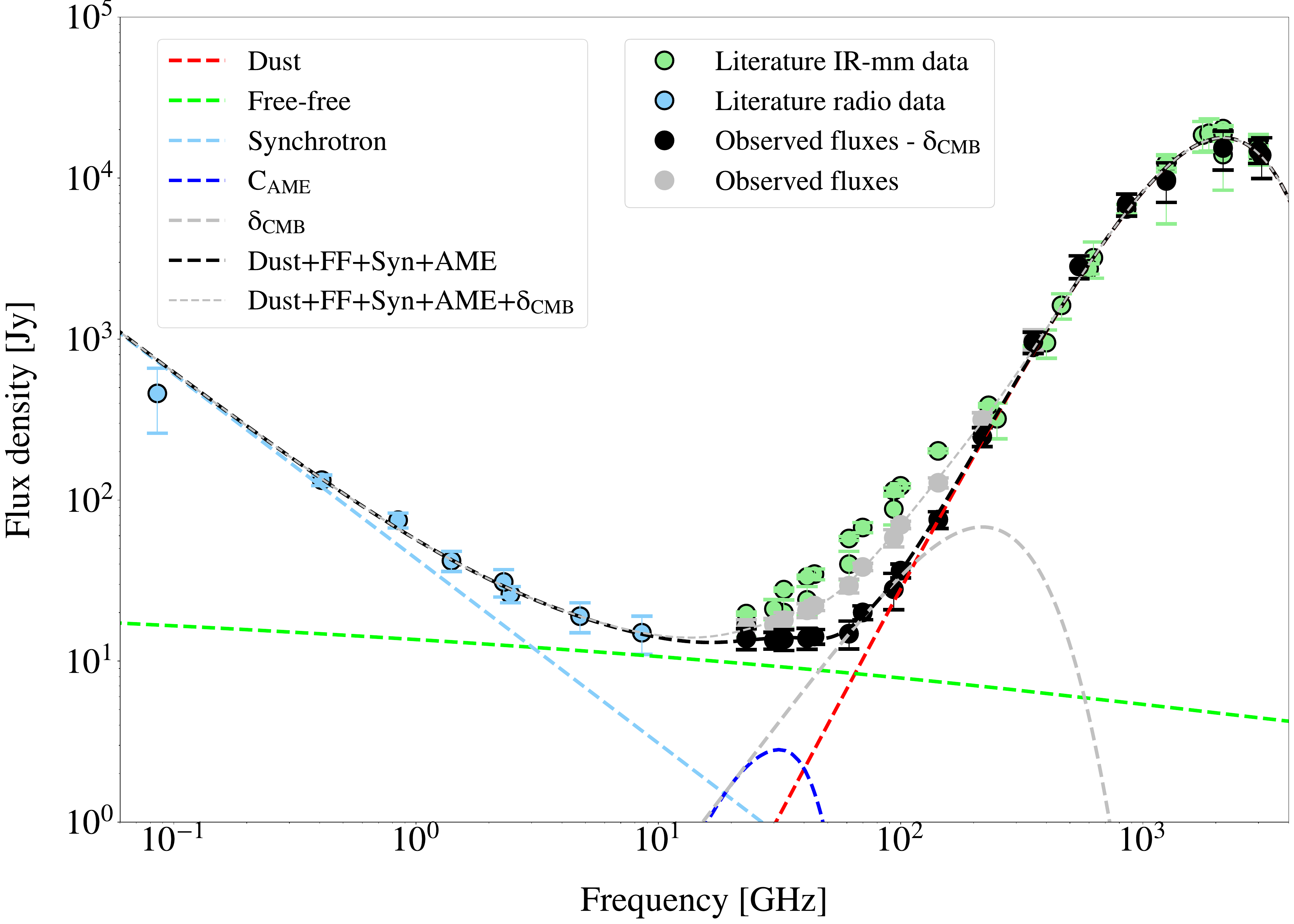}\hfill\includegraphics[width=0.5\textwidth]{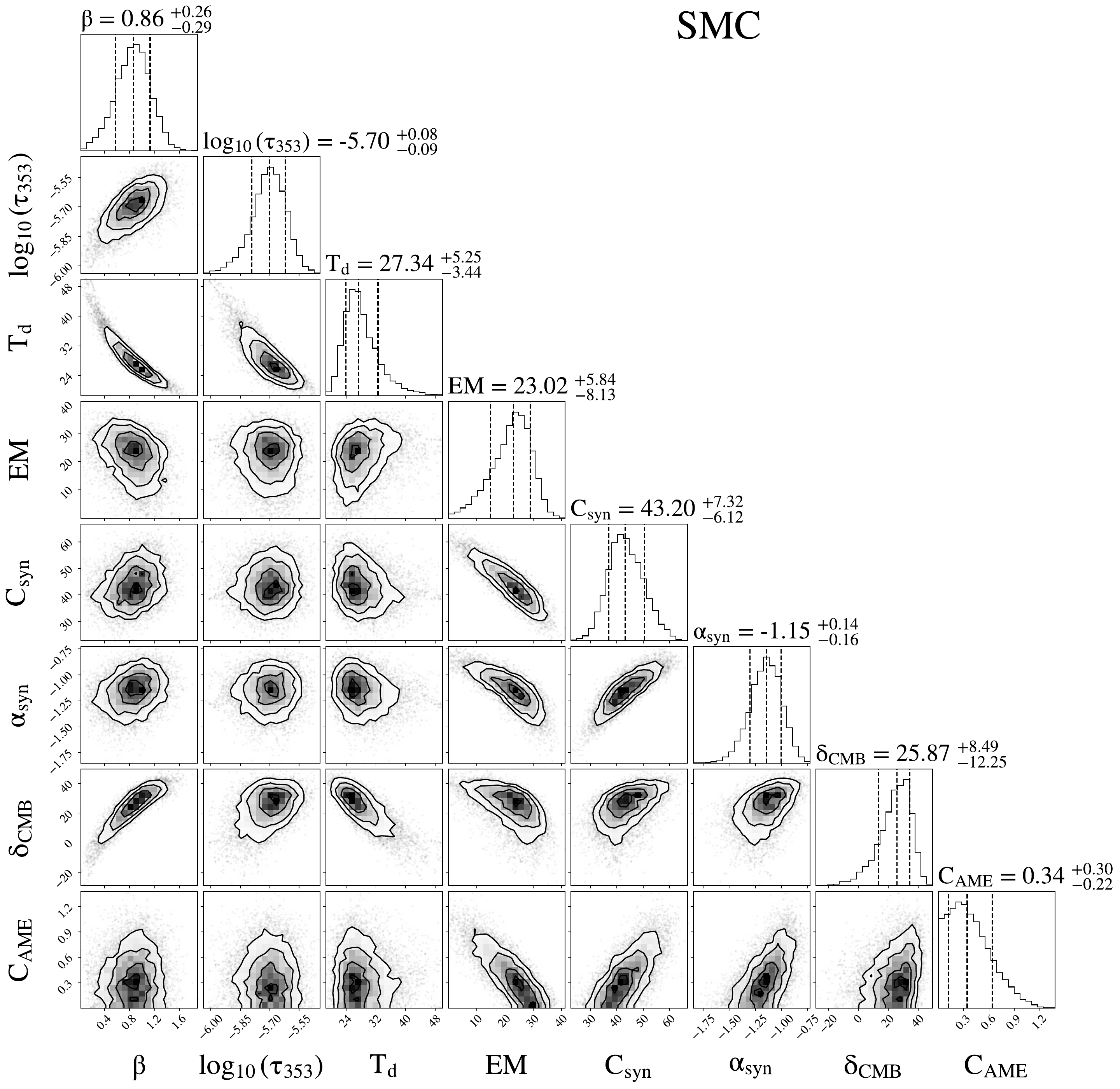}
    \caption{SEDs of LMC (top) and SMC (bottom) as observed with gray points and without CMB with black points (subtracted from the best model), and radio data in light blue (from Tables \ref{table_radio_data_lmc} for LMC and \ref{table_radio_data_smc} for SMC). Data points from the literature \citep{israel_2010,planck_2011} are overlaid in green. The best fit model spectra are overlaid for the global model and individual emission components. On the right, the corner plot displays probability distributions of each free parameter of the models. The values of the best-fit parameters obtained with the MCMC are written above each histogram.}
    \label{seds_lmc_smc}
\end{figure*}

\begin{figure*}
    \centering
    \includegraphics[width=0.5\textwidth]{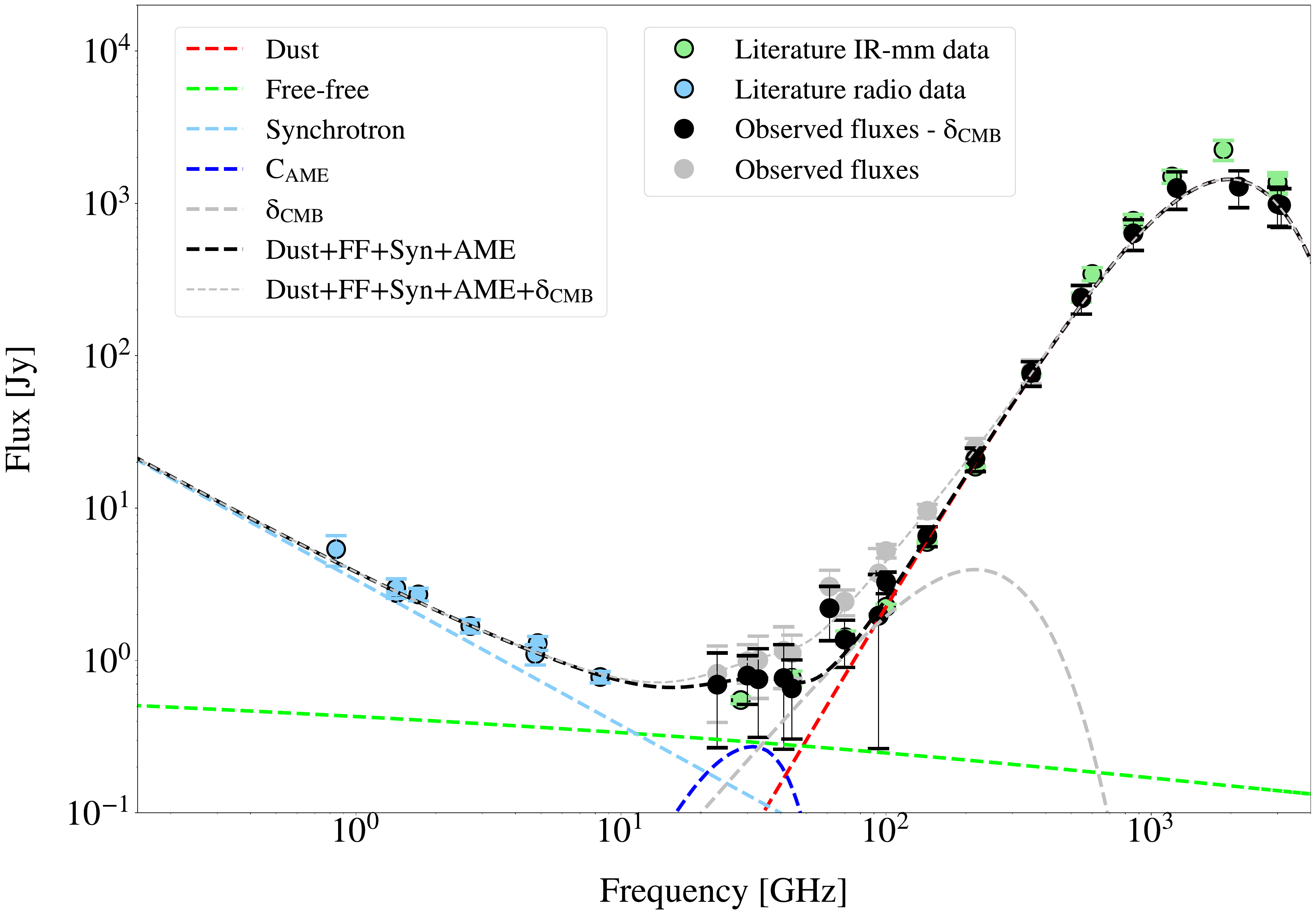}\hfill\includegraphics[width=0.5\textwidth]{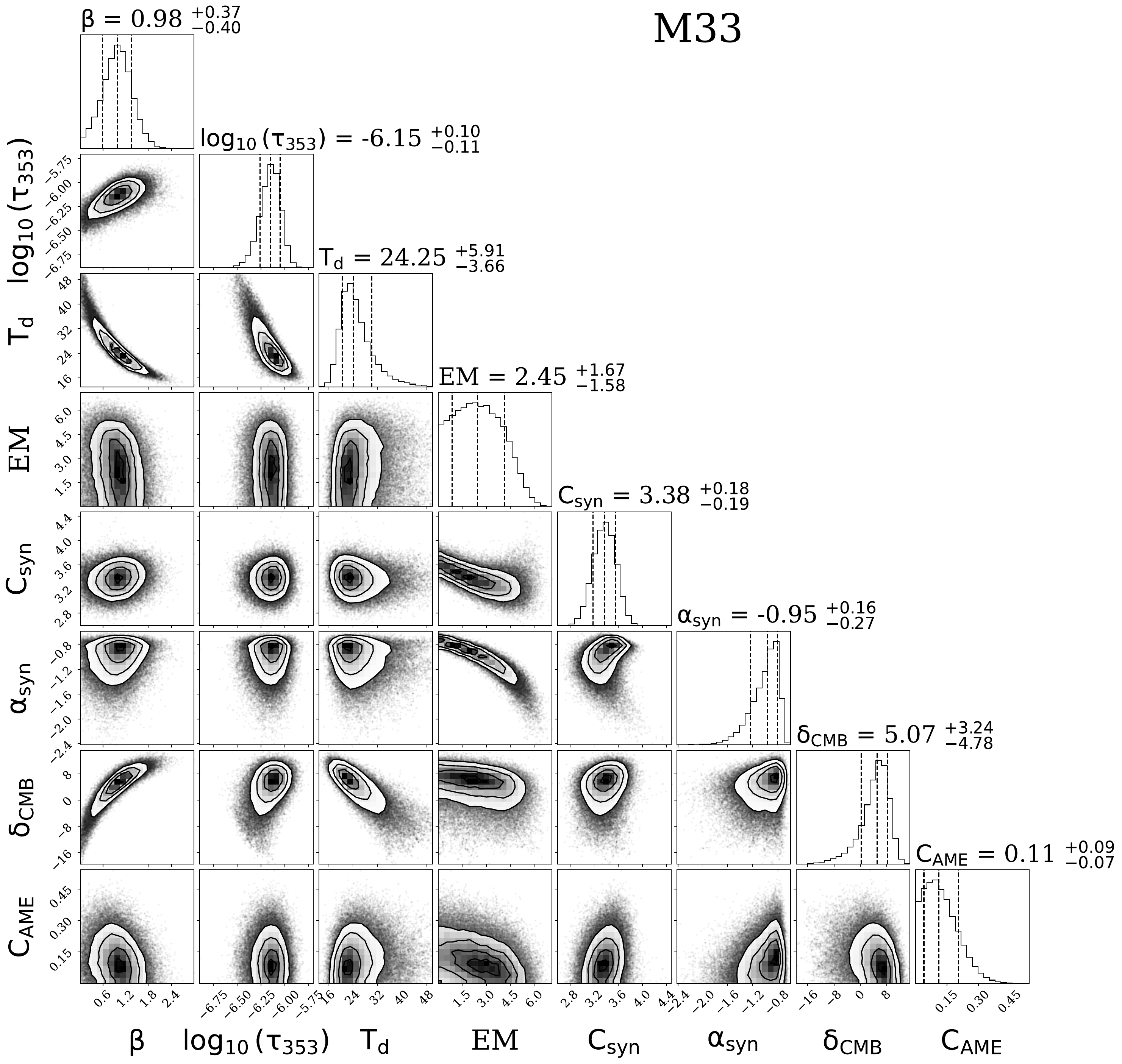}
       \caption{Same as Figure \ref{seds_lmc_smc} for M33. IR literature data are overlaid in green \citep{tibbs}, and radio data in light blue (from Table \ref{table_radio_data_m33}).}
    \label{seds_m33}
\end{figure*}

\begin{figure*}
    \centering
    \includegraphics[width=0.5\textwidth]{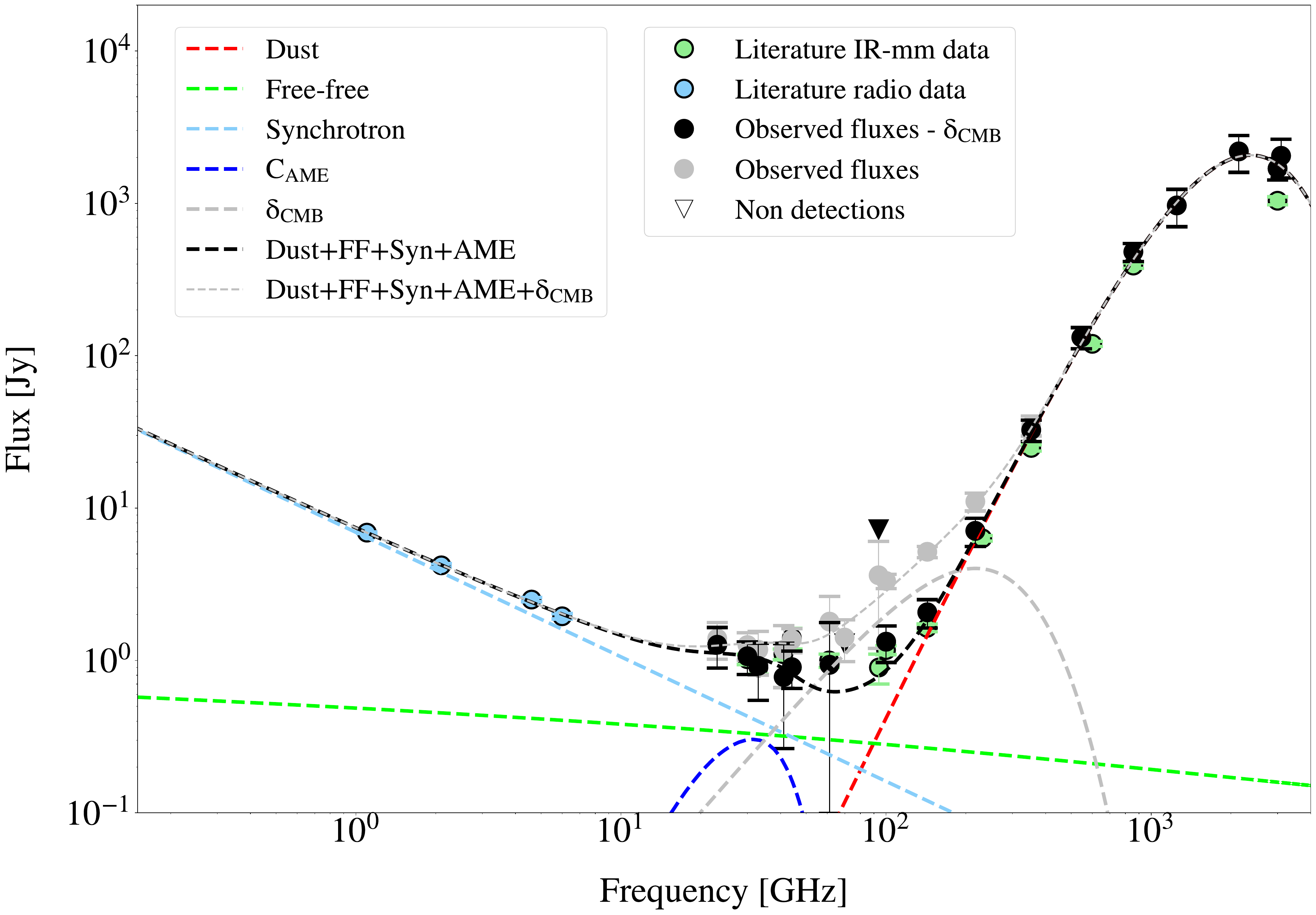}\hfill\includegraphics[width=0.5\textwidth]{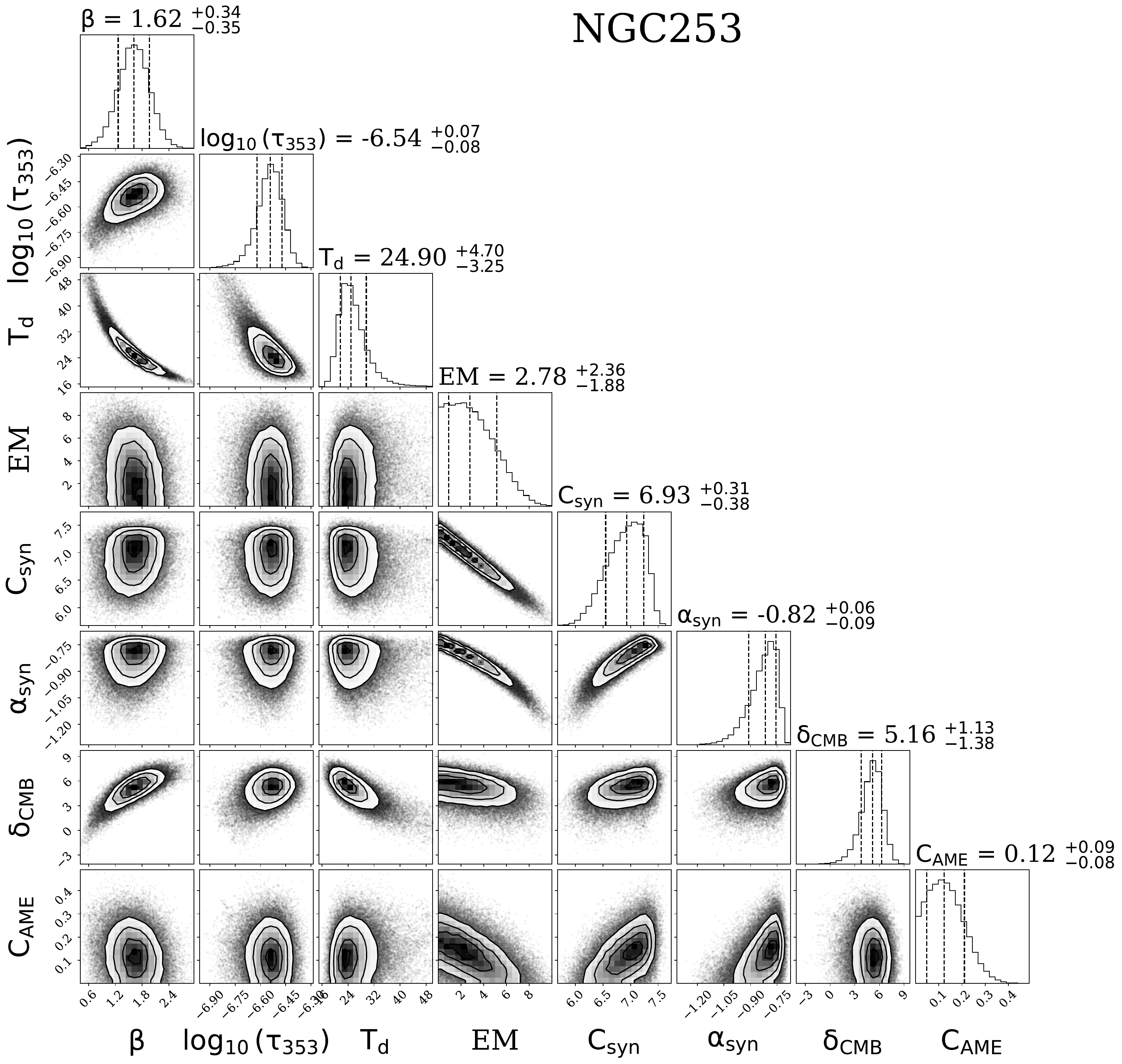}
    \includegraphics[width=0.5\textwidth]{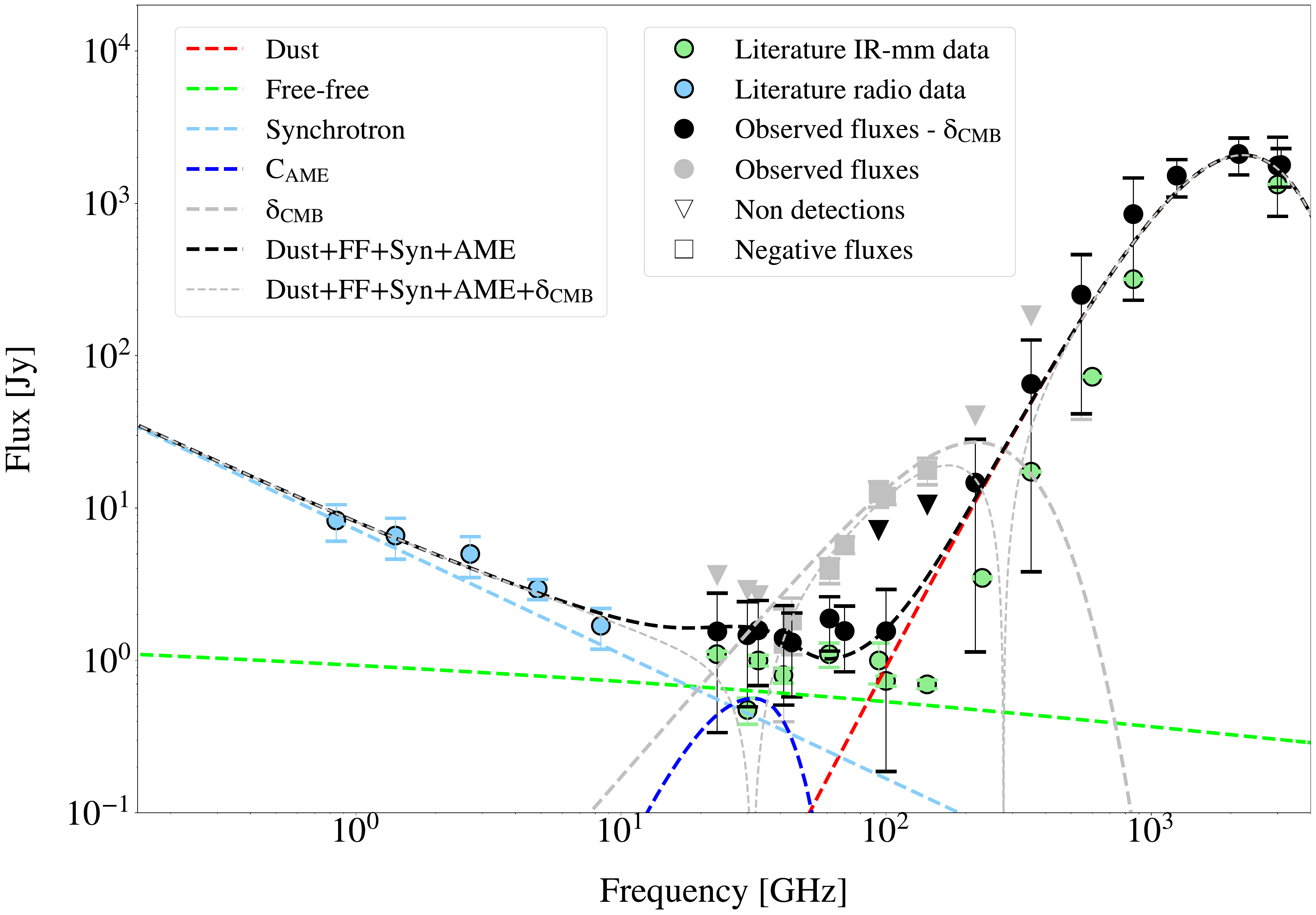}\hfill\includegraphics[width=0.5\textwidth]{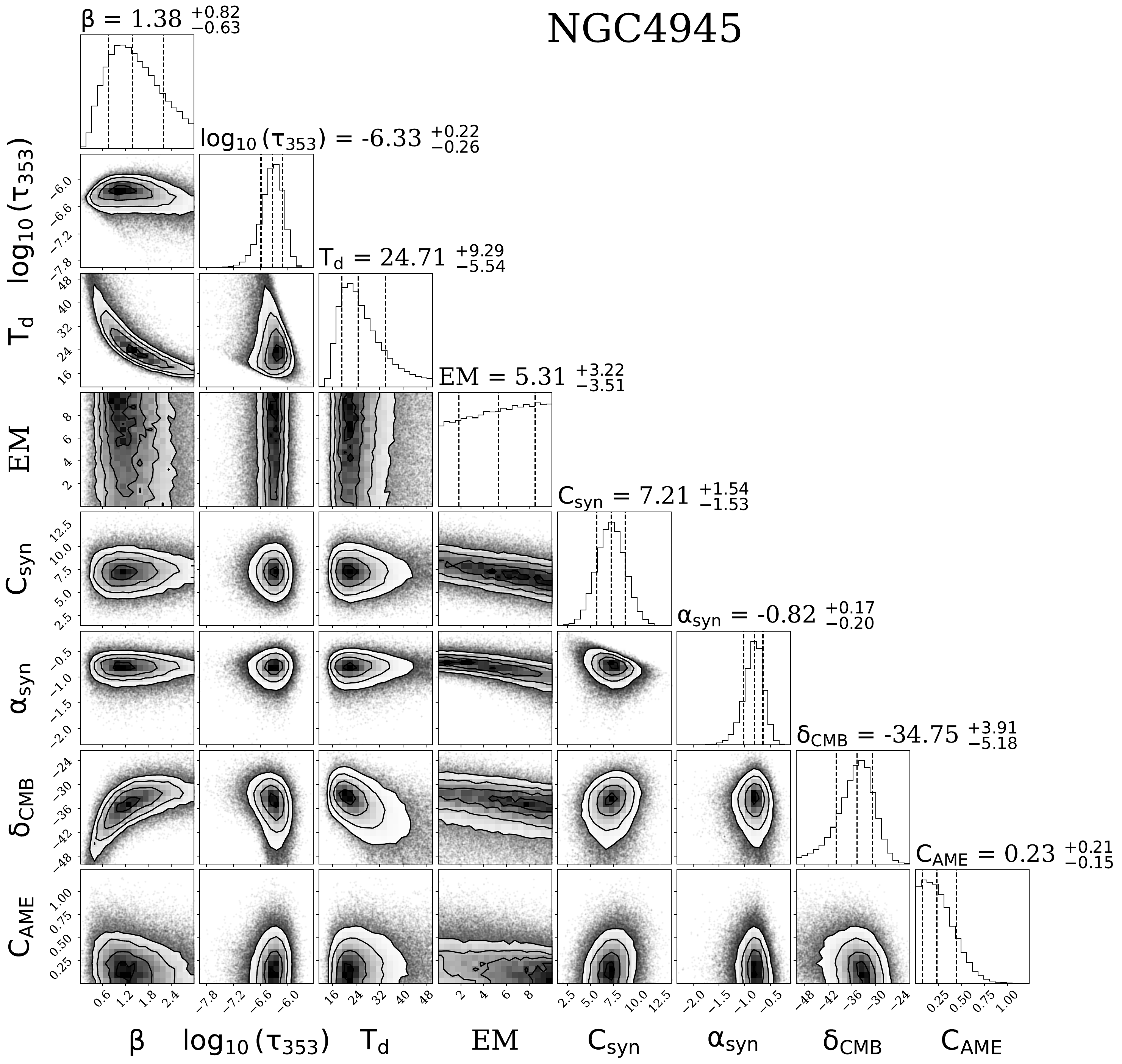}
    \caption{Same as Figure \ref{seds_lmc_smc} for NGC~253 (top), and NGC~4945 (bottom). Negative values are plotted as squares, using the absolute values of the measured flux densities. Non-detections are represented by upside-down triangles. IR literature data are overlaid in green \citep{dustpedia}, and radio data in light blue (from Tables \ref{table_radio_data_ngc253} for NGC~253 and \ref{table_radio_data_ngc4945} for NGC~4945).}
    \label{seds_ngc253_ngc4945}
\end{figure*}

We present the results for each galaxy in the sample: M31, LMC, SMC, M33, NGC~253, and NGC~4945. The following subsections describe the parameters derived for each galaxy, with the observed integrated SEDs and the corresponding best-fit models (Figures \ref{sed_m31}, \ref{seds_lmc_smc}, \ref{seds_m33}, and \ref{seds_ngc253_ngc4945}), together with the corner plots showing the parameter probability distributions. Best-fit parameters are listed in Table \ref{parameters_combined}. 

\subsection{M31}\label{results_m31}

M31 is a well-detected galaxy across all photometric bands, has been extensively studied even in the mm--cm domain, and is the only nearby galaxy where the AME was detected on the integrated SED \citep{planck2015,battistelli2019,harper_2023,fernandez_torreiro_m31}. We therefore use the results obtained for this galaxy as a detailed example and to check the reliability and advantages of our methodology. 

\begin{figure}[!ht]
   \centering
   \includegraphics[width=0.5\textwidth]{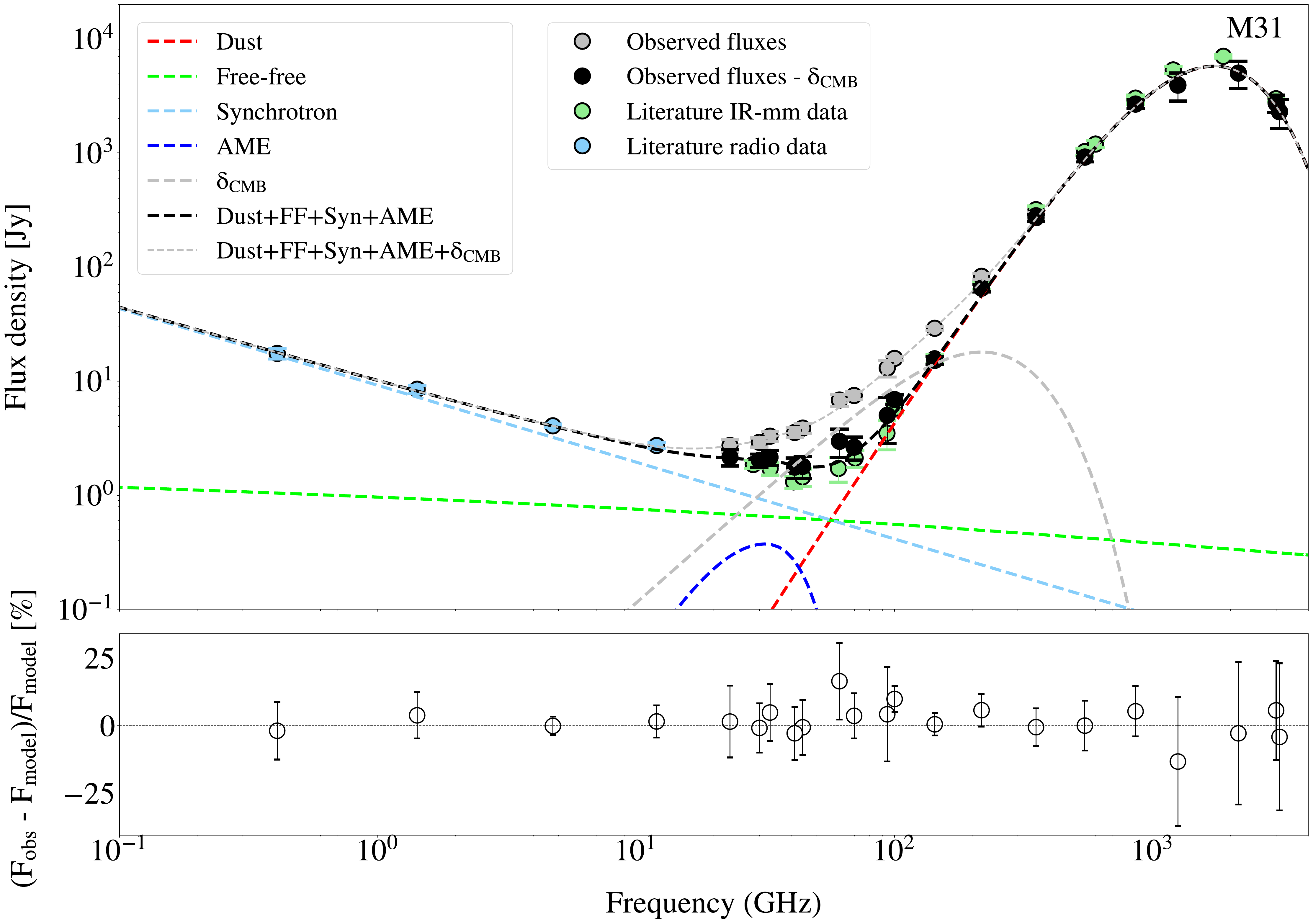}
   \caption{Integrated SED of M31. The gray points represent the flux densities of the galaxy with the CMB fluctuations in the background. This is the SED that is fitted with our model, together with the radio data from the literature (blue points). The black points correspond to the flux densities of the galaxy where the CMB fluctuations are subtracted with the values obtained with the best model of $\mathrm{\delta_{CMB}}$. The different emission components of our model obtained for the best fit are also represented. The blue points correspond to radio data from the literature. The green points correspond to literature data from \citet{Fritz_2012,Bennett_2013,planck2015}. On the bottom panel, residuals between the observed and modeled flux densities of M31 are shown.}
   \label{sed_m31}
\end{figure}

The observed SED integrated in the region of M31 is presented in Figure \ref{sed_m31} in gray points. The curves corresponding to the best fit model with the individual emission components are overlaid (best fit parameters are listed in Table \ref{parameters_combined}). Because our interest lies in the emission from the galaxy itself, we removed the $\delta_{\rm CMB}$ fluxes obtained from the best fit model and overplot the resulting SED of the galaxy alone with black points. 

\begin{figure}[!ht]
   \centering
   \includegraphics[width=0.5\textwidth]{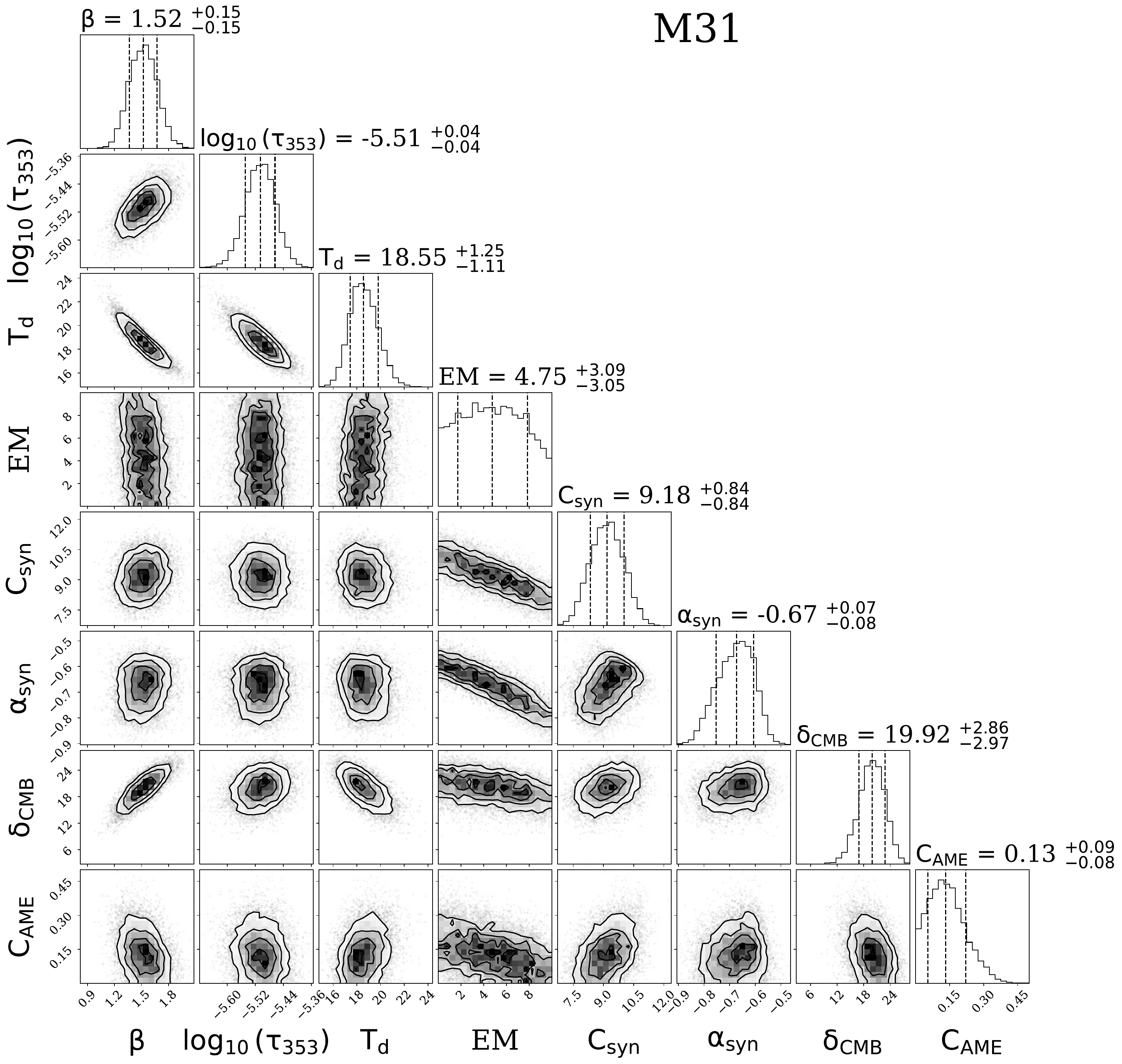}
   \caption{Posterior probability distributions for each free parameter of the model for the SED of M31. The values of the best-fit parameters obtained with the MCMC are written above each histogram.}
   \phantomsection\label{cornerplot_m31}
\end{figure}

As seen in the right panel of Figure \ref{sed_m31}, our simple model is enough to reproduce the emission observed within the uncertainties. With the data set we have, we find that we do not need any break in the dust emissivity index and a simple modified blackbody is enough to reproduce the thermal dust emission. This could result from the large spatial scales examined and the large uncertainties in the DIRBE fluxes, which influence the shape of the IR peak. The best fit parameters we obtain for dust emission correspond to an emissivity index $\beta$ = 1.52 $\pm$ 0.15 and a dust temperature $T_{d}$ of 18.55 $\pm$ 1.17 K. This is consistent with results obtained previously \citep{planck2015,fernandez_torreiro_m31,harper_2023}\footnote{\citet{planck2015} found $\beta$ = 1.62 $\pm$ 0.11 and $T_{d}$ = 18.2 $\pm$ 1.0 K;  \citet{fernandez_torreiro_m31} deduced $\beta$ = 1.71 $\pm$ 0.08 and $T_{d}$ = 18.49 $\pm$ 0.79 K; \citet{harper_2023} obtained $\beta$ = 1.44 $\pm$ 0.09 and $T_{d}$ = 19.15 $\pm$ 0.8 K}.

Analyzing the probability distributions of our model parameters (Figure \ref{cornerplot_m31}), we first observe the well-known and strong anti-correlation between the thermal dust parameters $\beta$ and $T_{d}$ due to the degeneracies between them in the simple modified blackbody model for thermal dust \citep{Dupac2003,Shetty_2009, ysard_2012}. This degeneracy is amplified by the uncertainties we have in all bands. Most of these uncertainties are mainly driven by imperfections in the foreground and background subtractions and are known to be correlated between bands. Taking the correlation of uncertainties into account could reduce the degeneracies between our dust parameters \citep{Gordon}. This is not the main focus of this study and is postponed to further work. We do however keep the full probability distribution of $\beta$ and $T_{d}$ for each galaxy when comparing these parameters between galaxies, as these degeneracies are distinct yet superimposed on actual $T_{d}$ - $\beta$ variations (see Section \ref{sec:dustprop}).

We also observe a clear degeneracy between $\beta$ and $\delta_{\rm CMB}$. This is expected as $\beta$ drives the slope of the long wavelength dust emission and an increase of positive $\delta_{\rm CMB}$ in the background of a galaxy may mimic a flatter spectral shape in the millimeter regime. This is exactly what we observe in the $\delta_{\rm CMB}$-$\beta$ joined probability distribution for M31. Within the uncertainties, the SED of M31 can be reproduced similarly with a steeper dust emissivity and a higher CMB fluctuation in the background, or a shallower dust emissivity and a lower positive CMB fluctuation in the background of this galaxy. For M31, flux densities are however sufficiently bright that both parameters can be detected significantly despite this correlated uncertainty.

Finally, we observe a degeneracy between the free-free and synchrotron parameters, hindering a precise determination of the free–free emission level in our analysis. As we can see in Figure \ref{cornerplot_m31}, the uncertainty on the free-free parameter EM is very large and we only measure an upper limit. The synchrotron spectral index we obtain ($\upalpha_{\rm syn}$ = $-0.67 \pm 0.07$) is consistent with the one obtained by \citet{harper_2023} ($-0.66 \pm 0.03$) and \citet{fernandez_torreiro_m31} ($-0.97\pm0.21$), yet lower than values obtained by \citet{planck2015} ($-0.92 \pm 0.16$) or \citet{battistelli2019} ($-1.1\pm 0.1$). These differences can be attributed to the radio flux densities used in the different studies and in particular the change obtained with the C-BASS measurement \citep{harper_2023} that increased the M31 flux density at 4.76 GHz by a factor of $\sim 2$ by including extended radio emission. 

M31 is currently the only nearby galaxy where AME was detected in the integrated SED \citep{planck2015,battistelli2019,harper_2023,fernandez_torreiro_m31}. However, the significance of the detection and amplitude of the AME vary between studies, in particular in view of the change of radio flux densities with C-BASS observations as mentioned above, but also depending on the way CMB fluctuations were taken into account in the line-of-sight of this galaxy. In this context, \citet{harper_2023} investigated in detail how different treatments of the CMB affect the derived properties of M31 by testing several Planck CMB component-separation maps. Their analysis showed that the flux density associated with the CMB component is typically 0.2–0.3 Jy at 30 GHz, with a scatter of about 0.1 Jy, highlighting that the choice of CMB model or subtraction method can significantly influence the inferred amplitudes of the other emission components. For instance, \citet{planck2015} reported an AME amplitude of $0.7 \pm 0.3$ Jy at 30~GHz, \citet{battistelli2019}obtained $1.45 \pm 0.18$ Jy, \citet{harper_2023} derived $0.27 \pm 0.09$ Jy, and \citet{fernandez_torreiro_m31} found a greater AME amplitude using the SMICA map ($1.17 \pm 0.29$ Jy) compared to COMMANDER map ($0.92 \pm 0.34$ Jy), degrading from a 4 $\upsigma$ to a 2.7 $\upsigma$
detection. The larger discrepancies are largely due to differences in CMB treatment and in the inclusion of flux from large-scale measurements such as C-BASS
detection. 

In our analysis, using a consistent modeling approach where CMB fluctuations are fitted simultaneously with the galaxy emission components, we find an AME amplitude of $0.13 \pm 0.08$ Jy at 30~GHz. This corresponds to a detection significance of only $1.5\sigma$, i.e., AME is not significantly detected. Among all results, \citet{battistelli2019} and \citet{fernandez_torreiro_m31} report the highest AME amplitude. However, \citet{planck2015} and \citet{harper_2023} results are consistent with ours to better than $2\upsigma$, showing overall agreement when accounting for the uncertainties. In Figure \ref{cornerplot_m31}, we can observe that the strength of the AME in our modeling is slightly degenerate with the $\delta_{\rm CMB}$, but also with the free-free and the synchrotron emission. Globally, with our results, AME appears too faint to be detected in comparison to other components at the same wavelengths and our measured uncertainties.

This in depth analysis of M31 in comparison of previous studies supports our modeling choices and highlights the importance to be aware of degeneracies between model parameters, not only between the emission components ($\beta$-$T_{\rm d}$ or between the synchrotron and free-free emission) but also with the CMB fluctuations in the background.

\subsection{The Magellanic Clouds}
The SEDs of the Magellanic Clouds are well reproduced by our model, with dust emissivity indices of $\beta = 1.29 \pm 0.18$ for the LMC and a notably low $\beta = 0.86 \pm 0.27$ for the SMC. \citet{planck_2011} reported steeper values ($\beta = 1.48 \pm 0.25$ for the LMC and $\beta = 1.21 \pm 0.27$ for the SMC). In contrast, we find higher dust temperatures than those derived by \citet{planck_2011}$,$ who obtained $T_{\mathrm{d}} = 21.0 \pm 1.9$~K for the LMC and $T_{\mathrm{d}} = 22.3 \pm 2.3$~K for the SMC, while we derived $T_{\mathrm{d}} = 24.38 \pm 2.33$~K and $T_{\mathrm{d}} = 27.3 \pm 4.3$~K, respectively. For the SMC, the AME is only marginally detected, with $\mathrm{C_{AME}}$ = 0.34 $\pm$ 0.25 Jy, and not detected in the LMC with $\mathrm{C_{AME}}$ = 0.20 $\pm$ 0.21 Jy.

\subsection{M33}

We derived a particularly low $\beta$ = 0.98 $\pm$ 0.37 and a warmer $T_{d}$ of 24.25 K $\pm$ 4.69 compared to previous studies from \citet{tibbs} where CMB fluctuations were subtracted from the maps. For the radio emission, we derived $C_{\mathrm{syn}}$ = 3.45 $\pm$ 0.17 and $\upalpha_{\mathrm{syn}}$ = $-0.91 \pm 0.16$. In comparison, \citet{tibbs} constrained $C_{\mathrm{syn}}$ while allowing $\upalpha_{\mathrm{syn}}$ to vary freely, obtaining 
$C_{\mathrm{syn}}$ = $-1.03 \pm 0.03$, which is slightly steeper than the value derived in this work. In the mm, the AME detection is marginal, with $\mathrm{C_{AME}}$ = 0.11 $\pm$ 0.08 Jy, peaking at 31 GHz, indicating AME is a minor component at 1.4 $\upsigma$ in the integrated SED.

\subsection{NGC~253}
For NGC~253, both the dust spectral index and temperature are lower and warmer, respectively, than previously reported in \citet{peel}. We found $\beta$ = 1.61 $\pm$ 0.34 and $T_{d}$ = 24.90 K $\pm$ 3.97, while \citet{peel} derived $\beta$ = 1.96 $\pm$ 0.11 and $T_{d}$ = 22.6 K $\pm$ 1.3, but in the case where CMB
fluctuations were subtracted from the maps. For the radio emission, we derived $C_{\rm syn}$ = 6.93 $\pm$ 0.34 and $\alpha_{\rm syn}$ = $-0.81 \pm 0.07 $, whereas \citet{peel} reported $C_{\rm syn} = 11.1 \pm 4.3$ and $\alpha_{\rm syn}$ = $-1.59 \pm 0.35$. Our results indicate a lower synchrotron coefficient and a flatter spectral index, reflecting the differences in the datasets, frequency coverage, and the treatment of CMB and other components in the modeling. For the AME emission, while \citet{peel} found a 3~$\upsigma$ upper limit of 0.14 Jy, while we found $\mathrm{C_{AME}}$ = 0.12 $\pm$ 0.08 Jy peaking at 31 GHz.

\subsection{NGC~4945}
For NGC~4945, we obtained a dust spectral index and temperature of $\beta = 1.37 \pm 0.72$ and $T_{d} = 24.71 \pm 7.41$~K, respectively, which correspond to a lower $\beta$ and higher $T_{d}$ than the values reported by \citet{peel} ($\beta = 2.5 \pm 0.2$ and $T_{d} = 18.9 \pm 1.1$~K), derived from maps where CMB fluctuations had been subtracted. In the radio domain, our measurements indicate a lower synchrotron normalization and a flatter spectral index than those reported by \citet{peel}. We measured $C_{\rm syn} = 7.21 \pm 1.53$ and $\alpha_{\rm syn}$ = $-0.81 \pm 0.18$, while \citet{peel} reported $C_{\rm syn} = 12.3 \pm 3.1$ and $\alpha_{\rm syn}$ = $-1.15 \pm 0.20$. For the AME component, we found $\mathrm{C_{AME}} = 0.23 \pm 0.18$~Jy, peaking at 31~GHz, whereas \citet{peel} obtained only a 3~$\upsigma$ upper limit of 0.13~Jy.

\subsection{Comparing results for the whole sample of nearby galaxies}

Since we have analyzed our sample of 6 galaxies in a consistent way, we now describe results for all galaxies and what we can learn from the comparison between them.

\subsubsection{CMB fluctuations in the background of our sample of galaxies} \label{cmb_fluctuations}

We observe that for all our bright and nearby galaxies, the CMB fluctuations in the background are significant in the mm regime, accounting for 25.8\% to 88\% of the 100 GHz emission (galaxy + $\delta_{\rm CMB}$). The importance of this contribution is probably amplified by the 1$\degree$ resolution adopted for this study. Indeed, this 1$\degree$ scale corresponds also to the first and strongest peak seen in the power spectrum of the CMB temperature fluctuations. As highlighted by \citet{planck_2014}, even bright Galactic sources can be affected by CMB fluctuations, leading to a small but generally positive bias in measured fluxes, particularly at frequencies below 100~GHz. \citet{harper_2023} extended this analysis to the galaxy M31, showing that the CMB can contribute a significant fraction of the total flux at microwave frequencies, and that fitting the CMB spectrally is often highly degenerate with other emission components such as free-free, AME, and synchrotron emission. While this is unfortunate and prevents us from detecting a wider sample of very nearby galaxies, it also means that higher resolution studies with instruments like the Sardinia Radio Telescope \citep{SRT} or the Atacama Large Aperture Sub-mm/mm Telescope \citep{Klaassen_2020,Liu_2024} will bring more insight on the mm--cm emission of galaxies. Nevertheless, nearby galaxies can cover very large areas on the sky, so observations must still map sufficiently large regions to properly account for both the galaxy emission and the background. Failing to do so can lead to missing a fraction of the total flux, which is particularly challenging at radio frequencies where instrumental and atmospheric effects limit accurate measurements of extended emission.

More surprisingly than the significance of CMB fluctuations towards the line-of-sight of galaxies, what is striking with our study is the fact that our modeling implies a positive CMB fluctuation in the background of 5 out of the 6 galaxies. Such positive fluctuations in the background were already reported individually in M31 \citep{planck2015,fernandez_torreiro_m31,harper_2023} or in the LMC and SMC \citep{bot2010,planck_2011}. However, this is the first time that this is observed in a consistent way in a sample of nearby galaxies and the quasi systematic positive $\delta_{\rm CMB}$ are clearly revealed. 
In Section \ref{discussion}, we discuss the significance of this effect.

\subsubsection{The case of the SMC\label{sec:smc}}

For the SMC, we observe that the data points from previous studies \citep{israel_2010,planck_2011} are well above the gray points from our study, both being taken with CMB fluctuations kept in. The green points delineating a flat emissivity shape were the reasons for \citet{bot2010} to claim the existence of a microwave excess above the expected emission. In \citet{planck_2011}, this excess was partly attributed to a positive CMB fluctuation in the background of the SMC and fully attributed to a positive CMB fluctuation in the background of the LMC. When comparing flux densities from \citet{planck_2011} to the one we get (our black points), both with CMB fluctuations removed, both are however completely consistent despite the different methodology. We tracked down this puzzling inconsistency to the fact that our foreground subtraction is slightly different and that our background subtraction is done using maps where the CMB fluctuations have been removed. Surprisingly, in the annulus around the SMC where this background is estimated (both in \citealt{planck_2011} and in this work) several negative CMB fluctuations are present in the CMB fluctuation maps from any component separation we have checked (including the COMMANDER map we use to remove the CMB fluctuations before estimating the background flux). The strong strength of the initial microwave excess reported by \citet{bot2010} might hence be not only due to a positive CMB fluctuation in the background of the SMC but also the subtraction of negative CMB fluctuations in the surrounding of the SMC. It is worth noting that systematic colder CMB temperatures have been highlighted to appear in the surrounding of galaxies \citep{Hansen_2023,luparello_2022,lambas_2024}, potentially tracing an unaccounted CMB foreground that could also exist in the SMC as well. It is also possible that the scale of the SMC and of the background annulus correspond unfortunately to a peak and a trough of the CMB temperature fluctuation power spectrum, a situation that will tend to favor an anti-correlation between the CMB temperature fluctuations observed at these scales. This possibility was raised in \citet{Ferraro2015} who computed a formalism to estimate the impact of CMB fluctuations in aperture photometric measurements. Analyzing the occurrence of CMB fluctuations in the background or surrounding of galaxies in a larger sample is therefore required to go further and is discussed in the context of our small sample in section \ref{cmb_fluctuations}.

\subsubsection{Dust properties}\label{sec:dustprop}

The dust thermal emission parameters we obtain for our sample of galaxies show large variations. While M31 is well described with dust emission with an emissivity index $\beta =1.52\pm 0.15$ similar to our Milky Way \citep[1.60 $\pm$ 0.06;][]{planck_xiv_2014}, M33 and SMC are distinguished by a particularly low $\beta$ of $0.98\pm 0.37$ and $0.86\pm 0.27$ respectively. Even accounting for the large uncertainties, these values are much lower than the dust emissivity index observed for thermal dust emission in our solar neighborhood.
Although slightly higher, we also find a low $\beta$ of $1.29\pm 0.18$ for the LMC. Such low emissivity index (even to just phenomenologically describe dust emission) in galaxies can be due to a combination of effects: 1) mixing of temperatures with larger amounts of colder dust  \citep{galliano_2003,paradis_2009}; 2) different dust grain composition \citep{Paradis_2010}; 3) a flattening of the intrinsic dust emissivity \citep{Coupeaud_2011} or effects in amorphous solids \citep{meny2007}; or 4) magnetic dipole emission at long wavelengths \citep{draine_2012}. These low $\beta$ values are the same phenomenology as large sub-millimeter excess as reported in galaxies \citep[e.g.,][]{galametz_2011,remy-ruyer-2013,galliano2022} obtained with models that include a fixed and shallower emissivity index (e.g., $\beta=[1.5-2]$). These low-$\beta$, sub-millimeter excess or flat dust emission have been reported already in M33 and SMC \citep{galliano_2003,bot2010,planck_2011,paradis2012,galliano_2018,tibbs} and are more widely observed in dwarf galaxies, low metallicities or regions of low densities \citep{remy-ruyer-2013,galliano_2018}. More recent work by \citet{Paradis_2024} also reports similarly low values of $\beta$ ( = 1, and defined as the slope between 850~$\upmu$m and 1.38~mm), in Galactic and nearby galaxy environments, from neural-network-based predictions of dust emission maps at submm–mm wavelengths, further supporting the picture of a flat dust emission spectrum at long wavelengths. Given the $\beta$-$\delta_{\rm CMB}$ degeneracy discussed in section \ref{cmb_fluctuations}, getting higher $\beta$ values with our model would be possible but would imply even higher positive $\delta_{\rm CMB}$ than currently reported.

\begin{figure*}[!ht]
   \centering
   \includegraphics[width=0.8\textwidth]{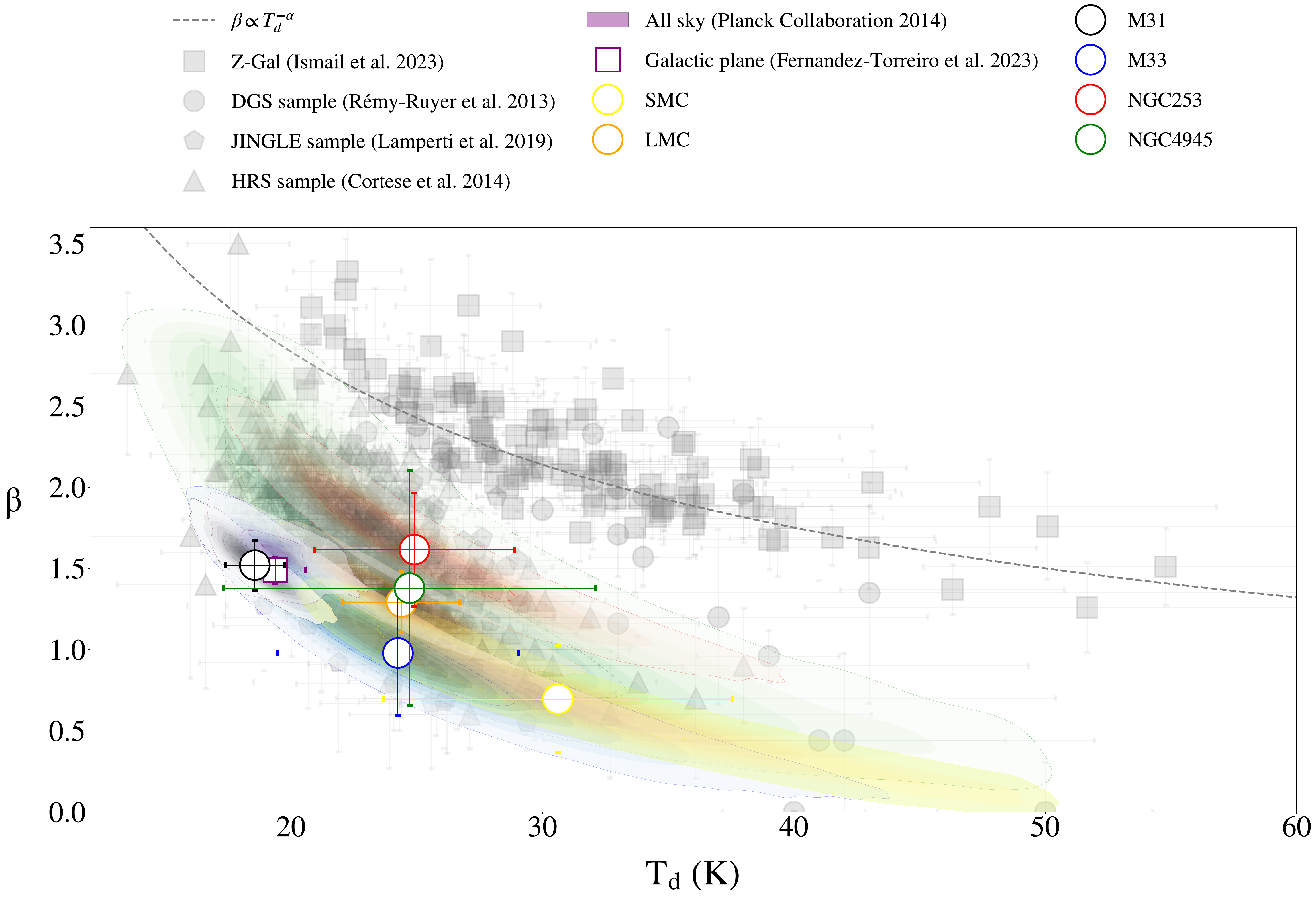}
   \caption{Comparison between the dust parameters $\beta$ and $T_{d}$ for the 6 galaxies studied in this work. The variations observed are compared to the same parameters obtained for other environments: the all-sky survey of \textit{Planck} \citep{planck_xi_2014}, the Milky Way \citep{fernandez-torreiro}, dwarf galaxies \citep{remy-ruyer-2013}, nearby galaxies from the \textit{Herschel} Reference Survey \citet{cortese_2014} and for high-redshift galaxies from \citet{ismail_2023}.}
   \label{temp_beta}
\end{figure*}

Degeneracies between $\beta$ and $T_{d}$ are clearly seen in all probability distributions obtained with our modeling (see Figures \ref{cornerplot_m31}, \ref{seds_lmc_smc}, \ref{seds_m33}, and \ref{seds_ngc253_ngc4945}). Despite these, intrinsic dust property changes are also observed beyond these degeneracies. We compare in Figure \ref{temp_beta} our best fit $\beta$ and $\rm T_{d}$ values for the 6 nearby galaxies analyzed in this work. 
Despite the intrinsic degeneracies as witnessed by the probability distributions, we observe that the $\beta$ values deduced are significantly different between our 6 galaxies, hence showing actual variations of the dust thermal emission slope between these galaxies. We observe a trend of lower $\beta$ values for higher dust temperatures, even if this trend is difficult to disentangle from the degeneracy between the two parameters. This anti-correlation has been widely reported in both observational and theoretical works \citep[e.g.,][]{Desert_2008, Paradis_2010, Smith_2012, Juvela_2013, Kirkpatrick_2014, cortese_2014}, and more recently in high-redshift studies such as in \citet{ismail_2023}. This $\beta$ - $T_{d}$ relation could reflect both the intrinsic properties of interstellar dust grains and the physical characteristics of the environments in which they reside. Such an inverse dependency has also been observed in laboratory experiments of interstellar grain analogs, including both amorphous carbons and silicates \citep{Agladze_1996, Mennella_1998,Demyk_2017}.

For comparison, we include in Figure \ref{temp_beta} results from other studies in different galaxy samples: dwarf galaxies from the DGS sample \citep{remy-ruyer-2013}, more massive systems from the \textit{Herschel} Reference Survey \citep{cortese_2014}, main sequences galaxies from the JINGLE survey \citep{Lamperti_2019}, high-redshift galaxies from \citet{ismail_2023}, and for the Galactic plane from \citet{fernandez-torreiro}.
The results we obtained for our sample are globally consistent with the general trends observed at higher resolution with other local galaxies, including the Milky Way. As reported in \citet{ismail_2023}, galaxies at high redshift display a similar decreasing $\beta$ with increasing $T_{d}$ yet is shifted to systematically higher $\beta$ for a given dust temperature. In our sample, we observe that the starburst galaxy NGC~253 also appear shifted (although less than high-z galaxies) with respect to local group galaxies in our sample (LMC, SMC, M31, M33). All galaxies are however broadly compatible with other galaxy samples in our nearby universe \citep{remy-ruyer-2013,cortese_2014,Lamperti_2019}. These differences in the long wavelength shape of dust in galaxies could be related to differences in ISM conditions, such as variations in radiation field intensity, dust grain growth, or changes in dust composition. A lower $\beta$ might indicate larger or more processed grains, while higher temperatures in high-redshift galaxies could be associated with stronger star formation activity and a more intense interstellar radiation field. Whatever the physical processes at stake, these results confirm that dust properties depend on the environment and vary between galaxies.

\subsubsection{Anomalous Microwave Emission}

The detection of AME in M33, NGC~253, NGC~4945, and SMC is marginal. Our best fit parameters for the AME flux densities are summarized in Table \ref{parameters_combined} and the probability distribution functions obtained are visible in Figures \ref{cornerplot_m31}, \ref{seds_lmc_smc}, \ref{seds_m33}, and \ref{seds_ngc253_ngc4945}. We only found a small correlation between the AME parameter, ${C_{\rm AME}}$, and the free-free parameter EM, and not a visible correlation with the other parameters, meaning that AME is relatively independent of the other components, yet it seems to be a very minor component of the integrated emission of the galaxy. Comparing the models with and without the AME component (values summarized in Table \ref{parameters_combined}), the parameters change only modestly and remain consistent within their uncertainties. The inclusion of AME results in only minor reductions in $\chi^{2}$ for few galaxies (SMC, M31, and NGC~253), while in some galaxies (LMC, M33, and NGC~4945) the fit is even slightly degraded. Overall, these results indicate that AME is a minor contributor to the integrated emission. Since AME has been more clearly detected in individual well resolved extragalactic regions \citep{murphy2010,hensley2015,murphy2018}, this systematic non-detection of AME in all galaxies of our sample could be due to the fact that it is only a significant process in specific regions of galaxies and not visible in integrated SED due to the dilution of this component on large-scales. 

One can wonder whether our non-detection are consistent with what would be expected from the dust emission observed in these galaxies. To do this, we express our AME best-fit components into AME emissivities. The AME emissivity is defined in \citet{fernandez-torreiro} as the ratio between the AME intensity at its peak frequency and the thermal dust intensity at 3000 GHz (100 $\upmu$m):
\begin{equation}
    \mathrm{\epsilon^{AME}_{peak}} = \frac{T^{\mathrm{AME}}_{\mathrm{peak}}}{I_{3000 \, \mathrm{GHz}}} = \frac{\frac{c^{2}}{2k_{B}\nu^{2}}\, S^{\mathrm{AME}}_{\mathrm{peak}}}{S_{3000\,\mathrm{GHz}}}
\end{equation}
We use the AME emissivity to normalize the AME by dust content, enabling meaningful comparison of its strength across regions with different dust amounts.

The AME emissivities computed from our best fits for our sample of galaxies are given in Table \ref{parameters_combined}.
Our derived upper limits on $\epsilon^{\rm AME}_{\rm peak}$ are lower than typical values observed in our Galaxy of 10 to 14 $\upmu$K/(MJy/sr) \citep{davies_2006,planck_xxiii_2015,Harper_2022,fernandez-torreiro}. This confirms that the AME component in our integrated sample of galaxies is weak compared to Galactic standards. To account for the sensitivity of 3000~GHz-based emissivities to dust temperature variations \citep[e.g.,][]{Tibbs_2012}, we also computed the AME emissivity normalized by the optical depth at 353~GHz $\tau_{353}$. Using this normalization, we obtain values of 5.94 $\pm$ 0.69~K/$\tau_{353}$ for M31, 34.13 $\pm$ 10.70~K/$\tau_{353}$ for M33, 86.67 $\pm$ 20.28~K/$\tau_{353}$ for NGC~253, 53.14 $\pm$ 33.20~K/$\tau_{353}$ for NGC~4945, 3.85 $\pm$ 0.64~K/$\tau_{353}$ for the LMC, and 13.72 $\pm$ 3.47~K/$\tau_{353}$ for the SMC. The difference between our 3000~GHz- and $\tau_{353}$-normalized emissivities reflects the sensitivity of the 3000~GHz emission to dust temperature. Even modest variations in dust temperature can strongly affect the measured brightness, leading to systematically lower and more scattered emissivities, while normalization by longer-wavelength tracers better accounts for residual temperature variations. These $\tau_{353}$ emissivity results suggest that AME, while present, is a minor contributor to the integrated emission of these galaxies, consistent with our earlier conclusion based on the marginal detections. However, the relatively higher emissivities in M33, NGC~253, and NGC~4945 may indicate that AME is more prominent in specific regions or environments with favorable physical conditions, such as dense gas or strong radiation fields, even if the integrated galaxy signal remains low. These findings align with Galactic studies and extragalactic work, highlighting that AME is strongest in localized environments and can be diluted when averaging over whole galaxies \citep{Tibbs_2012,planck_xi_2014,dickinson_2018,bianchi_2022}, although our measured value for M31 is higher than the one reported by \citet{harper_2023}. AME is probably not a dominant emission component at the scale of a whole galaxy and may only be significant in localized environments where physical conditions favor spinning dust emission. Since specific physical conditions are required to excite efficiently \citep{Draine_1998_b}, it might not be surprising that spinning dust emission would be concentrated in dense regions or environments where the radiation field is intense \citep{tibbs_2013, hermelo2016}. Thermal dust emission at high frequencies reflects the overall dust distribution heated by ambient radiation, while AME originates from specific grain populations localized in certain regions. As a result, averaging AME emissivity over large areas dilutes the signal due to regions with little or no AME, producing our observed low average values. Although detecting AME in the integrated spectra of galaxies remains challenging, it can serve as a tracer of the relative contribution of dense regions within galaxies. In particular, AME detections have been reported in some galaxies where conditions favor significant small grain rotation, such as a star-forming region in NGC 6946 (first extragalactic detection; \citet{murphy2010,hensley2015} and a compact radio source associated with NGC 4725 \citep{murphy2018}. Since higher resolution observations or redshifted SEDs of more distant galaxies will be less impacted by the CMB fluctuations in the background, there is an interest in continuing the search for extragalactic AME, either in resolved regions within nearby galaxies or in integrated SEDs of more distant galaxies.

\subsubsection{Free-free emission}

In our model, the free-free emission is parametrized with the emission measure EM. The best fit parameters obtained for all galaxies are summarized in Table \ref{parameters_combined}. Except for the SMC where free-free emission is clearly detected, we find that the integrated SEDs of our sample of galaxies are reproduced with sub-dominant free-free emission. EM is often poorly constrained in our study. The large uncertainties stem both from the degeneracies between the synchrotron, AME emissions and CMB anisotropies, as we can see in Figures \ref{cornerplot_m31}, \ref{seds_lmc_smc}, \ref{seds_m33}, and \ref{seds_ngc253_ngc4945}, and from the relatively low contribution of the galaxy flux densities at long wavelengths, where the synchrotron and CMB components dominate. The inverse correlation we observe in the probability distribution between the synchrotron strength and slope and the free-free emission strength could potentially be solved with priors on the physical processes that drive an actual correlation between these two components. Indeed, thermal and non-thermal emission processes can be associated with star formation in galaxies and known to be connected on large scales \citep[e.g.,][]{tabatabaei_2013_m31,tabatabaei_2013_ngc6946}. However, the strength of these physical correlations are non-trivial and can vary depending on the spatial scale and the physical conditions of the interstellar medium and the structure of magnetic fields.


\section{Discussion}\label{discussion}

\subsection{Assessing the impact of CMB fluctuations} \label{cmb_fluctuations_discussion}

Our modeling infers a positive $\mathrm{\delta_{CMB}}$ in 5 out of 6 galaxies which is surprising. The significance of this quasi-systematic positive $\mathrm{\delta_{CMB}}$ in the background of our sample of galaxies can be examined by estimating the probability of observing at least 5 positive fluctuations among 6 galaxies using a binomial distribution. Under the assumption that CMB temperature fluctuations are randomly distributed following a symmetric Gaussian distribution centered on zero, the resulting probability we compute is $\approx 9.4 \%$, which is not formally significant. However, we also need to take into account the amplitudes of the $\mathrm{\delta_{CMB}}$ we infer from our modeling, as several are substantially larger than their associated uncertainties. The mean of the 6 $\mathrm{\delta_{CMB}}$ inferred values is 11.79~$\upmu$K, which is inconsistent at 3.78 $\sigma$ with a null expectation when the individual errors are taken into account. From that perspective, the $\mathrm{\delta_{CMB}}$ in the background of our sample of galaxies are therefore very significative. 

However, CMB temperature fluctuations have very specific spatial distributions that are witnessed as clear peaks in the CMB fluctuation power spectrum. In order to evaluate the statistical relevance of our $\mathrm{\delta_{CMB}}$, with respect to the expected variations of $\mathrm{\delta_{CMB}}$ on the sky, we performed an MCMC analysis in which the CMB fluctuations were sampled at 6 random positions on the sky, using the same apertures as described in Section \ref{sec:projandaper}. For each realization, the mean flux density over the 6 apertures was computed, and the procedure was repeated 1000 times. The resulting distribution of mean values is centered on zero with a standard deviation of $\approx$ 55~$\upmu$K, which is much larger than our average <$\mathrm{\delta_{CMB}}$> $\approx$ 12 $\upmu$K. Although the fact that 5 out of 6 individual values are positive may appear noteworthy, especially given the significance of the $\mathrm{\delta_{CMB}}$ values we infer from our modeling, in the context of the CMB variance such a configuration remains probable at these angular scales. This analysis indicates that the observed mean is not statistically distinguishable from what would be expected due to random CMB fluctuations. 

Unexpected $\mathrm{\delta_{CMB}}$ effects have been reported in larger galaxy samples \citep{luparello_2022,Hansen_2023,lambas_2024}. Put in this context, our quasi systematics positive $\mathrm{\delta_{CMB}}$ might trace a similar effect in local galaxies. Yet, our positive $\mathrm{\delta_{CMB}}$ are currently not significant given the CMB variance to confirm and disprove such effect in nearby galaxies. To go further, it will hence require to extend the study to a much larger sample of galaxies, and/or data at higher resolution.

The case of the SMC where negative fluctuations of CMB temperatures are present in the surroundings, while a positive CMB temperature fluctuation is inferred from our modeling of the observed emission towards the line-of-sight of the SMC itself, adds to this questioning. The low $\beta$ for the dust emission and the $\beta$-$\mathrm{\delta_{CMB}}$ degeneracy also need to be better understood to improve our knowledge on these effect. More work is clearly needed especially at high angular resolution where we can expect intrinsically lower fluctuations of the CMB temperature. This would help to understand both this potential new CMB foreground component and the dust emission at long wavelengths.

\subsection{Variability of the AME peak frequency}

It is well established that in Galactic PDRs, the AME peak can shift toward mm wavelengths, \citep[e.g.,][]{Casassus_2008,Tibbs_2010,Tibbs_2011,Scaife_2010,Bell_2019,Cepeda_2021} where the peak position is found to vary with local physical conditions \citep{Ysard_2022}. Moreover, \citet{Poojon_2024} found that AME was detected in NGC 2903 and marginally in NGC 2146, with the spinning dust emission peaking at higher frequencies and showing stronger emissivity than previously reported. This supports the spinning dust model prediction that AME peaks shift to higher frequencies in denser environments, such as molecular clouds and PDRs. These results indicate that the AME peak frequency is not fixed, and raises the possibility that part of the mm excess observed in galaxies could be related to a shifted AME peak. However, the spectral shape of the excess emission we observe in this study is already well reproduced by a combination of thermal dust emission, CMB fluctuations, free–free, and synchrotron emission. The CMB component is particularly difficult to reconcile with a shifted AME spectrum. Testing such a scenario would require introducing additional free parameters, which would further increase degeneracies in the modeling. Indeed, studies that have attempted this approach often find very large shifts of the AME peak into the radio domain, raising questions about the physical plausibility of such solutions \citep{Poojon_2024}.

\subsubsection{Limitations of the radio data}\label{sec:limits_radio}

In our analysis, we incorporate radio data compiled from the literature. A fully consistent treatment would require processing the radio measurements in the same way as the FIR–cm data to ensure homogeneity across the entire wavelength range; however, this is beyond the scope of this study. While this is a weak point of our analysis, we believe this difference does not have an impact on the results. The use of radio data is needed, as these points provide constraints on the levels of free–free and synchrotron emission. Yet, the estimates do not need to be as accurate as in the FIR-mm to constrain the dust emission and $\mathrm{\delta_{CMB}}$.
We perform tests in which we artificially increased the uncertainties of the radio data points. While this increase changed the free-free and synchrotron inferred values, it does not impact our conclusions on dust and $\mathrm{\delta_{CMB}}$. We present the case of the LMC analysis with increased radio uncertainties in Appendix \ref{appendix_test_radio_data}. The fits indicate an increase in the free–free emission component EM, whereas the contribution from $\mathrm{\delta_{CMB}}$ shows only a marginal change. These tests confirm that our conclusions on CMB fluctuations are not solely driven by the uncertainties or potential biases in the radio measurements. Although the radio data available in the literature may carry significant uncertainties, they do not dominate the overall results of our fits.


\section{Conclusion}\label{conclusion}

We performed the first homogeneous analysis of 6 very nearby galaxies with an excellent wavelength coverage in the mm–cm domain. We used data from COBE-DIRBE, IRAS, \textit{Planck} and WMAP from 100 $\mathrm{\upmu}$m (3000 GHz) to 13 mm (22.8 GHz) at $1^{\circ}$ resolution to study 6 nearby galaxies: M31, M33, NGC~253, NGC~4945, LMC, and SMC. We find that the global SED can be well represented by a model of dust, free-free, synchrotron, and AME emissions, with CMB temperature fluctuations in the background. 

Our resulting parameters for the thermal dust emission, free-free and synchrotron are consistent with what has been seen in individual studies of these galaxies and previous results from the literature. We do highlight the very low dust emissivity indices that are inferred to describe the dust emission of the SMC, M33 or the LMC, noting that these $\beta$ values are purely phenomenological and do not reflect the actual intrinsic $\beta$ values of dust in these galaxies. Comparing our resulting dust parameters between galaxies, we do observe significant variations of $\beta$ and $T_{d}$ between galaxies. These variations are consistent with the one observed in larger samples of galaxies and can be compared to higher redshift samples.

While AME is present in the best-fit models for all galaxies in our sample, the detections are marginal ($\leq$1$\upsigma$) in each case, with AME emissivities lower than typical Galactic standards. This shows that AME is a very minor component in the integrated spectra of these nearby galaxies, consistent with previous studies. In M31, we measure an AME amplitude of $0.13 \pm 0.08$ Jy at 30~GHz ($1.5\sigma$), which is lower than earlier estimates such as \citet{battistelli2019} but consistent within uncertainties with \citet{planck2015} and \citet{harper_2023}. Differences between studies largely reflect the inclusion of large-scale flux (e.g., C-BASS) and the treatment of CMB and free-free emission. Overall, these results confirm that AME is weak compared to other emission components.

We emphasize that CMB fluctuations contribute significantly to the mm–cm flux at 1$\degree$ resolution in all the galaxies we analyzed. Our results highlight a suspicious inferred presence of a quasi systematic positive CMB temperature fluctuation in the background of 5 out of 6 galaxies. Taking into account the CMB variance, this could reflect a statistical bias due to the small size of our sample and the particular $1\degree$ scale of our study. Further work is required to check whether the $\mathrm{\delta_{CMB}}$ effect observe in the background of more distant galaxies could also exist in our studied local galaxies and trace unknown emission component that is not yet accounted for. Degeneracies between $\mathrm{\delta_{CMB}}$ and the dust emissivity index $\beta$ could play a role in changing the importance of these inferred positive CMB temperature fluctuations in the background of galaxies. However, we stress that decreasing $\mathrm{\delta_{CMB}}$ in the background would flatten the inferred dust emission shape at long wavelengths even more, a possibility that needs to be carefully assessed in light of current dust models.

Higher angular resolution studies will be needed to understand further both the long wavelength shape of dust emission in galaxies and use this knowledge to disentangle it from the CMB temperature fluctuation in the background of galaxies and their surroundings, including with a potential new foreground component. Current upgrades of the Sardinia Radio Telescope \citep{SRT}, further QUIJOTE observations \citep{QUIIJOTE2015} or the Atacama Large Aperture Submillimeter Telescope \citep{Klaassen_2020}, all hold the potential to study nearby galaxies in the mm--cm range with more details, hence reducing the uncertainty due to CMB temperature fluctuations and a better separation of foreground and background sources of emission that currently dominate our uncertainties. The advent of large radio facilities such as the Square Kilometer Array \citep[SKA][]{SKA2009} will drastically change our understanding of the thermal and non-thermal radio components in galaxies with unprecedented detail. Furthermore, galaxies at redshifts 2 and higher will have their rest-frame mm--cm emission observed with SKA-mid and could be detected, therefore providing access to this long-wavelength thermal dust, AME and free-free emission part of the galaxies, with any contributions from the CMB temperature fluctuations. These future facilities therefore hold strong promises for our understanding of the mm--cm emission of galaxies.

\section*{Data availability}
Maps of LMC, SMC, M33, NGC~253, and NGC~4945 before and after the subtraction of the foreground and background sources are available on the public platform \href{https://doi.org/10.5281/zenodo.17662333}{Zenodo}.

\begin{acknowledgements}
We would like to thank M.-A. Miville-Deschênes and B. Hensley for insightful discussions at different stages of this work. We thank the referee for useful comments which helped improve the content of the paper.
\end{acknowledgements}


\bibliographystyle{aa} 
\bibliography{biblio}


\begin{appendix}

\section{Radio data} \label{sec:radiodata}

\begin{table}[!ht]
\caption{Radio data of LMC.}      
\label{table_radio_data_lmc}      
\centering          
\begin{tabular}{ccc}      
\hline 
\vspace{0.2cm}
& References & \\
Wavelength & Frequency & $\mathrm{F_{\upnu}}$ \\ 
(cm) & (GHz) & (Jy) \\  
\hline
\vspace{0.2cm}
 & \citet{klein_1989} & \\
13.03 & 2.30 & 404 $\pm$ 30 \\
21.41 & 1.40 & 529 $\pm$ 30 \\
73.47 & 0.40 & 925 $\pm$ 30 \\
189.74\footnote{Mills (1959) but revised and listed by \citet{klein_1989} } & 0.15 & 1736 $\pm$ 490 \\\vspace{0.2cm}
 & \citet{haynes_1991} & \\
3.5 & 8.55 & 270 $\pm$ 35 \\
6.31 & 4.75 & 363 $\pm$ 30 \\
\hline                
\end{tabular}
\end{table}

\begin{table}[!ht]
\caption{Radio data of SMC.} 
\label{table_radio_data_smc}      
\centering          
\begin{tabular}{ccc}      
\hline 
\vspace{0.2cm}
& References & \\
Wavelength & Frequency & $\mathrm{F_{\upnu}}$ \\ 
(cm) & (GHz) & (Jy) \\  
\hline
\vspace{0.1 cm}
 & \citet{haynes_1991} & \\
3.5 & 8.55 & 15 $\pm$ 4 \\
6.31 & 4.75 & 19 $\pm$ 4 \\
12.23 & 2.45 & 26 $\pm$ 3 \\
74.94\footnote{Haslam (1992) but revised and listed by \citet{haynes_1991}} & 0.40 & 133 $\pm$ 10 \\
\vspace{0.1 cm}
 & \citet{mountfort_1987} & \\
13.03 & 2.30 & 31 $\pm$ 6 \\
\vspace{0.1 cm}
 & \citet{loiseau_1987} & \\
21.41\footnote{revised by \citet{leroy_2007}} & 1.4 & 42 $\pm$ 6 \\
\vspace{0.1 cm}
 & \citet{ye_1991} & \\
35.68 & 0.84 & 75 $\pm$ 8 \\
\vspace{0.1 cm}
 & \citet{klein_1989} & \\
374.74\footnote{Mills (1959) but revised and listed by \citet{klein_1989}} & 0.08 & 460 $\pm$ 200 \\
\vspace{0.1 cm}
 & \citet{alvarez_1989} & \\
749.48 & 0.04 & 415 $\pm$ 80 \\
\vspace{0.1 cm}
 & \citet{Shain_1959} & \\
1498.96 & 0.02 & 5270 $\pm$ 1054 \\
\hline                
\end{tabular}
\end{table}

\begin{table}[!ht]
\caption{Radio data of M31 \citep{harper_2023}.}      
\label{table_radio_data_m31}      
\centering          
\begin{tabular}{ccc}     
\hline                       
Wavelength & Frequency & $\mathrm{F_{\upnu}}$ \\
(cm) & (GHz) & (Jy) \\
\hline
2.5 & 12.0 & 2.73 $\pm$ 0.16 \\
6.3 & 4.76 & 4.06 $\pm$ 0.14 \\
21.1 & 1.42 & 8.50 $\pm$ 0.70 \\
73.5 & 0.41 & 17.50 $\pm$ 1.90 \\
\hline                
\end{tabular}
\end{table}

\begin{table}[!ht]
\caption{Radio data of M33.}      
\label{table_radio_data_m33}      
\centering          
\begin{tabular}{ccc}     
\hline 
\vspace{0.2cm}
& References & \\
Wavelength & Frequency & $\mathrm{F_{\upnu}}$ \\ 
(cm) & (GHz) & (Jy) \\  
\hline
\vspace{0.2cm}
 & \citet{Buczilowski} & \\
2.8 & 10.70 & 0.48 $\pm$ 0.15 \\
6.3 & 4.75 & 11.0 $\pm$ 0.167 \\        
11 & 2.70 & 1.68 $\pm$ 0.168 \\
17.4 & 1.72 & 2.71 $\pm$ 0.254 \\
21.1 & 1.42 & 2.99 $\pm$ 0.44 \\
\vspace{0.2cm}
35.6 & 0.84 & 5.37 $\pm$ 1.22 \\
\vspace{0.2cm}
 & \citet{tabatabaei_2007} & \\
3.6 & 8.35 & 0.78 $\pm$ 0.06  \\
6.2 & 4.85 & 1.30 $\pm$ 0.13 \\
21 & 1.42 & 2.76 $\pm$ 0.06 \\
\hline
\end{tabular}
\end{table}

\begin{table}[!ht]
\caption{Radio data of NGC~253 \citep{radio_data_m82}.}      
\label{table_radio_data_ngc253}      
\centering          
\begin{tabular}{ccc}      
\hline                       
Wavelength & Frequency & $\mathrm{F_{\upnu}}$ \\
(cm) & (GHz) & (Jy) \\
\hline
5.0 & 6.0 & 1.95 $\pm$ 0.08 \\
6.51 & 4.6 & 2.51 $\pm$ 0.08 \\
14.28 & 2.1 & 4.21 $\pm$ 0.12 \\
27.25 & 1.1 & 6.90 $\pm$ 0.08 \\
\hline                
\end{tabular}
\end{table}

\begin{table}[!ht]
\caption{Radio data of NGC~4945 \citep{radio_data_m82}.}      
\label{table_radio_data_ngc4945}      
\centering          
\begin{tabular}{ccc}      
\hline                       
Wavelength & Frequency & $\mathrm{F_{\upnu}}$ \\
(cm) & (GHz) & (Jy) \\
\hline
3.57 & 8.4 & 1.69 $\pm$ 0.5 \\
6.18 & 4.85 & 2.95 $\pm$ 0.45 \\
11.1 & 2.7 & 5.0 $\pm$ 1.5 \\
21.26 & 1.41 & 6.6 $\pm$ 1.98 \\
35.6 & 0.84 & 8.3 $\pm$ 2.24 \\
\hline                
\end{tabular}
\end{table}

\newpage
\section{Local estimate of the cirrus foreground}\label{appendix_dust_emission}

Around each galaxy we study, we use the region outside the galaxy as defined in Section \ref{sec:projandaper} to compare the global IR-mm emission and gas templates. We perform a linear fit to the correlations obtained on independent pixels between the observed emission in each band and the gas column density. We do this on the maps at 1$\degree$ resolution with the CMB fluctuations, CIB and resolved sources subtracted.

Since foreground cirrus at high Galactic latitude is dominated by diffuse neutral gas \citep{Boulanger_1988,boulanger_1996,planckxvii_2014},we use the HI 4-PI HI column density maps. We fit the observed dust-HI correlation as observed in each band, in independent pixels, in the surroundings of each galaxy. The average IR-HI emissivity ratios from \citet{planckxvii_2014} are used as initial guess for the fit.

Dust emission is associated to all gas phases and the contribution of dust associated to molecular or ionized gas can sometimes be significant, even at high Galactic latitude \citep[e.g.,][]{lagache1999,Cheng_2025}. We therefore checked whether adding the ionized gas using \citet{finkbeiner} H$\alpha$ map and the CO gas using the \citet{planckco} maps as additional gas tracers would improve the foreground removal. We considered the dust-HI, dust-${\rm H\alpha}$ and dust-CO relationships as 3D or 4D dust-gas correlation fits. In the surroundings of the galaxies we study, we found that the contributions of the dust correlations with H${\alpha}$ and CO were non significant. Since our goal is to subtract the emission foreground in front of the galaxies we study, we therefore restricted ourselves to the use of the dust-HI correlations around each galaxy.

The correlation between dust emission and HI column densities is fitted as a linear relationship outside each galaxy, in each band from 97 $\mathrm{\upmu}$m to 13 mm:  
\begin{equation}
    \mathrm{I_{\nu}} = \sigma_{\text{HI}} \times \, N_{\text{HI}} + y_{0}
\end{equation}
where:  
$I_{\nu}$ is the intensity map at frequency $\nu$, $N_{\text{HI}}$ the HI column density in units of $10^{20}~\rm  H/cm^{2}$, $\sigma_{HI}$ represents the dust emissivity per unit HI column density and $y_{0}$ accounts for any residual background offset in the emission. We determined the best-fit values using a linear least-squares regression technique.

\begin{figure}[!h]
    \centering
    \includegraphics[width=0.9\columnwidth]{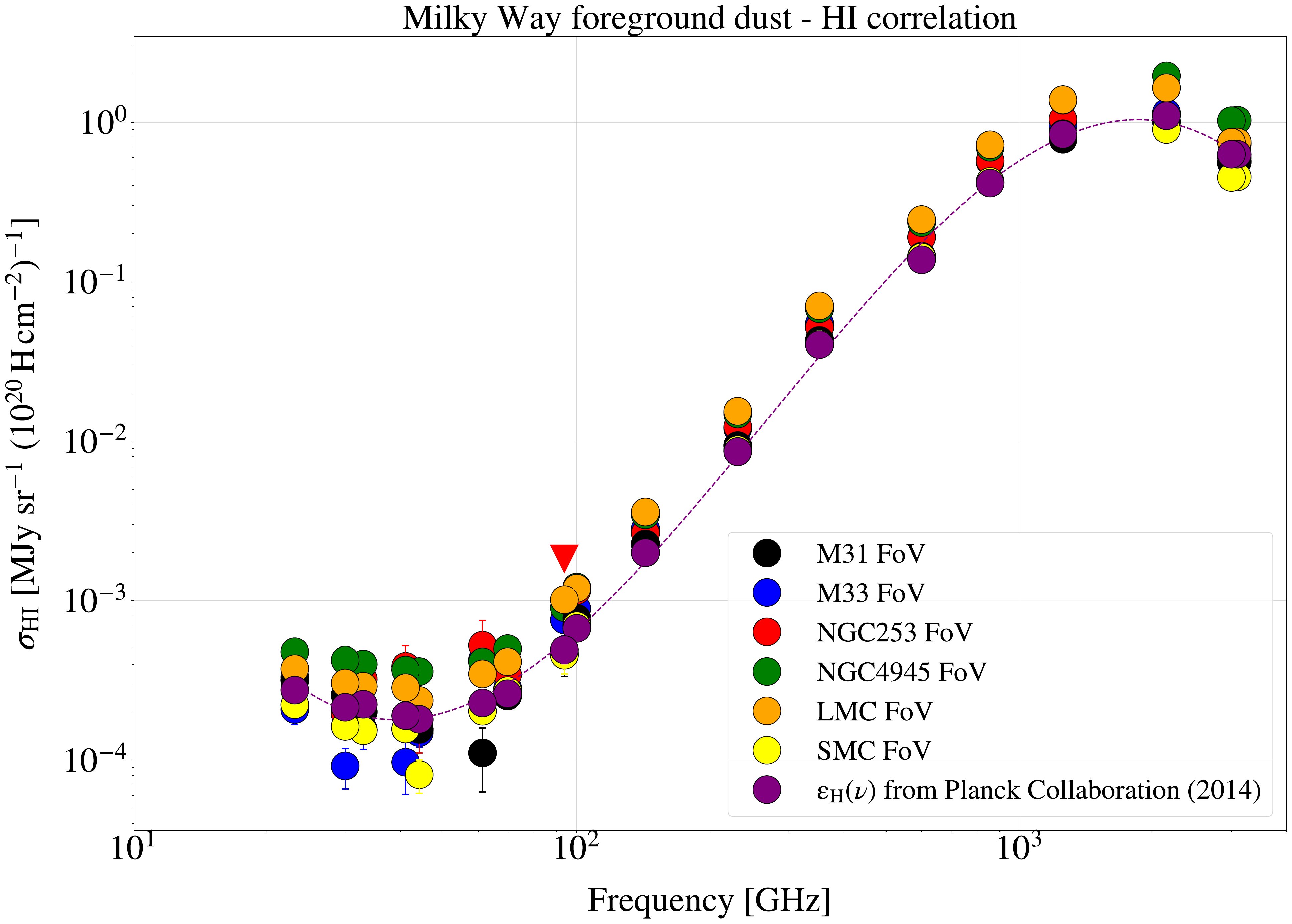}
    \caption{SEDs of the Milky Way foreground estimated from local dust-N$\mathrm{_{HI}}$ correlations around our 6 studied galaxies. These SEDs laws are compared to the average high-latitude values \citep{planckxvii_2014}. Non-detections are represented by upside-down triangles.}
    \label{sed_correlations_hi}
\end{figure}

The dust emissivities per HI column that we obtained in each band are plotted as a function of the observed wavelength in Figure \ref{sed_correlations_hi} for each region around the studied galaxies separately. We observe that the Milky Way dust SEDs that we deduce are globally consistent across the whole wavelength range with what was observed on average at high Galactic latitude by \citet{planck_2011diffuse}.  Yet, we also observe differences between regions which justify our use of the locally estimated ratios rather than the average high-latitude sky values. We apply these local emissivity coefficients to the HI4PI N$_{\rm HI}$ maps to create templates of the Milky Way foreground emission. Towards the line-of-sight of M31, part of the HI in the disk of M31 has velocities that intercept our velocity range $[-90,90]$~km/s and we want to avoid subtracting these as Milky Way dust. This is the only case where we observe this and hence we decided to mask this very specific region and to interpolate the HI maps in the direction of M31. Using our obtained dust emission templates for the Milky Way foreground, we subtract these maps at all wavelengths to our convolved and (CIB+sources+$\delta_{\rm CMB}$) subtracted maps for each galaxy of our sample.

\section{Best-fit parameters with and without AME component}\label{appendix_best_fit_parameters}

\begin{sidewaystable}
    \caption{Best–fit parameters obtained for each galaxy, using an all-components model (thermal dust, free–free, synchrotron, and AME emission for the galaxy as well as CMB temperature fluctuations in the background) and the model where the AME component is excluded. For the models including AME, the AME emissivities deduced from the AME best fit parameter, both normalized by the 3000~GHz flux density and by the dust optical depth $\tau_{353}$, are also added. \label{parameters_combined}}      
    \scriptsize
    \centering          
    \begin{tabular}{|c|*{6}{c}|*{6}{c}|}    
    \hline  
    Parameters & & & & All components model & & & & & & No AME model & & \\
     & LMC & SMC & M31 & M33 & NGC~253 & NGC~4945 & LMC & SMC & M31 & M33 & NGC~253 & NGC~4945 \\
    \hline
    $\beta$ & 1.29 $\pm$ 0.18 & 0.86 $\pm$ 0.27 & 1.52 $\pm$ 0.15 & 0.98 $\pm$ 0.37 & 1.61 $\pm$ 0.34 & 1.37 $\pm$ 0.72 & 1.28 $\pm$ 0.18 & 0.84 $\pm$ 0.28 & 1.59 $\pm$ 0.14 & 1.12 $\pm$ 0.37 & 1.68 $\pm$ 0.31 & 1.49 $\pm$ 0.72 \\  
    $\mathrm{log_{10}}(\tau_{353})$ & $-5.19 \pm 0.05 $ & $-5.70 \pm 0.08 $ & $-5.50 \pm 0.04 $ & $-6.15 \pm  0.10 $  & $-6.54 \pm 0.07 $ & $-6.33 \pm 0.23 $ & $-5.19 \pm 0.06 $ & $-5.69 \pm 0.08 $ & $-5.50 \pm 0.04$ & $-6.14 \pm 0.10$ & $-6.53 \pm 0.06$ & $-6.35 \pm 0.24$ \\
    $T_{d}$ (K) & 24.4 $\pm$ 2.33 & 27.3 $\pm$ 4.34 & 18.5 $\pm$ 1.17 & 24.2 $\pm$ 4.69 & 24.9 $\pm$ 3.97 & 24.7 $\pm$ 7.41 & 24.4 $\pm$ 2.31 & 27.6 $\pm$ 4.74 & 18.1 $\pm$ 1.07 & 22.9 $\pm$ 4.21 & 24.1 $\pm$ 3.38  & 23.9 $\pm$ 6.94 \\
    EM ($\mathrm{pc\,cm^{-6}}$) & 27.9 $\pm$ 23.0 & 23.02 $\pm$ 6.98 & 4.75 $\pm$ 3.06 & 1.81 $\pm$ 1.38 & 2.78 $\pm$ 2.12 & 5.31 $\pm$ 3.36 & 34.3 $\pm$ 24.9 & 30.1 $\pm$ 3.63 & 7.08 $\pm$ 2.59 & 3.34 $\pm$ 1.55 & 4.84 $\pm$ 1.99 & 6.21 $\pm$ 3.22 \\
    $C_{\mathrm{syn}}$ & 563 $\pm$ 47.1 & 43.2 $\pm$ 6.72 & 9.18 $\pm$ 0.83 & 3.45 $\pm$ 0.17 & 6.93 $\pm$ 0.34 & 7.21 $\pm$ 1.53 & 555 $\pm$ 51.0 & 37.3 $\pm$ 4.66 & 8.61 $\pm$ 0.71 & 3.29 $\pm$ 0.19 & 6.59 $\pm$ 0.32 & 7.00 $\pm$ 1.54\\
    $\upalpha_{\mathrm{syn}}$ & $-0.49 \pm 0.04$ & $-1.14 \pm 0.14$ & $-0.67 \pm 0.07$ & $-0.91 \pm 0.16$ & $-0.81 \pm 0.07$ & $-0.81 \pm 0.18$ & $-0.49 \pm 0.05$ & $-1.27 \pm 0.13$ & $-0.70 \pm 0.07$ & $-1.06 \pm 0.25$ & $-0.89 \pm 0.08$ & $-0.79 \pm 0.19$ \\
    $\mathrm{\delta_{CMB}}$ ($\mathrm{\upmu}$K) & 64.7 $\pm$ 11.7 & 25.8 $\pm$ 10.4 & 19.9 $\pm$ 2.91 & 5.36 $\pm$ 3.95 & 5.15 $\pm$ 1.25 & $-34.74 \pm 4.54$ & 62.5 $\pm$ 11.6 & 22.2 $\pm$ 10.2 & 21.4 $\pm$ 2.46 & 6.63 $\pm$ 3.29 & 5.21 $\pm$ 1.20 & $-33.58 \pm 4.01 $\\
    $\mathrm{C_{AME}}$ & 0.20 $\pm$ 0.21 & 0.34 $\pm$ 0.25 & 0.13 $\pm$ 0.08 & 0.11 $\pm$ 0.08 & 0.12 $\pm$ 0.08 & 0.23 $\pm$ 0.18 & - & - & - & - & - & - \\
    $\chi^{2}$ & 8.85 & 4.97 & 1.55 & 8.55 & 4.52 & 2.90 & 7.13 & 5.02 & 2.28 & 6.56 & 5.63 & 1.68 \\
    $\mathrm{\chi^{2}_{r}}$ & 0.80 & 0.38 & 0.16 & 0.57 & 0.45 & 0.6 & 0.59 & 0.36  & 0.21 & 0.44 & 0.51 & 0.14\\
    \hline
    $\mathrm{\upepsilon_{AME}}$ ($\upmu$K/(MJy/sr)) & 0.91 $\pm$ 0.97 & 5.96 $\pm$ 4.96 & 4.50 $\pm$ 2.94 & 9.00 $\pm$ 7.16 & 5.88 $\pm$ 3.98 & 10.47 $\pm$ 10.00 & - & - & - & - & - & - \\
    $\mathrm{\upepsilon_{AME}}$ (K/$\tau_{353}$) & 3.85 $\pm$ 0.64 & 13.72 $\pm$ 3.47 & 5.94 $\pm$ 0.69 & 34.13 $\pm$ 10.70 & 86.67 $\pm$ 20.28 & 53.13 $\pm$ 33.20 & - & - & - & - & - & - \\
    \hline              
    \end{tabular}
\end{sidewaystable}

\newpage
\onecolumn
\section{Test on limitation of the radio data}\label{appendix_test_radio_data}
\begin{figure*}[!h]
    \centering
    \includegraphics[width=0.5\columnwidth]{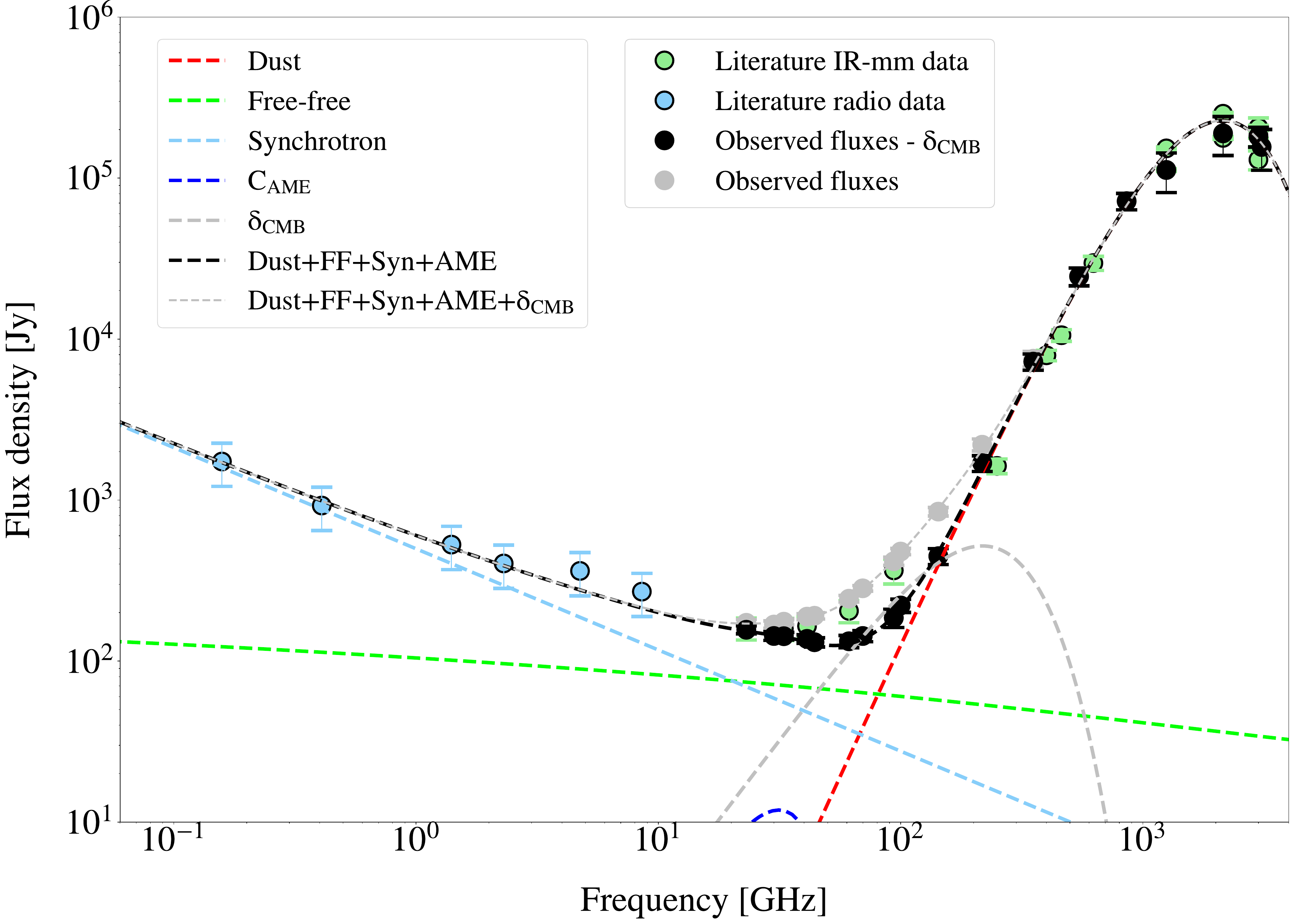}\hfill
    \includegraphics[width=0.44\columnwidth]{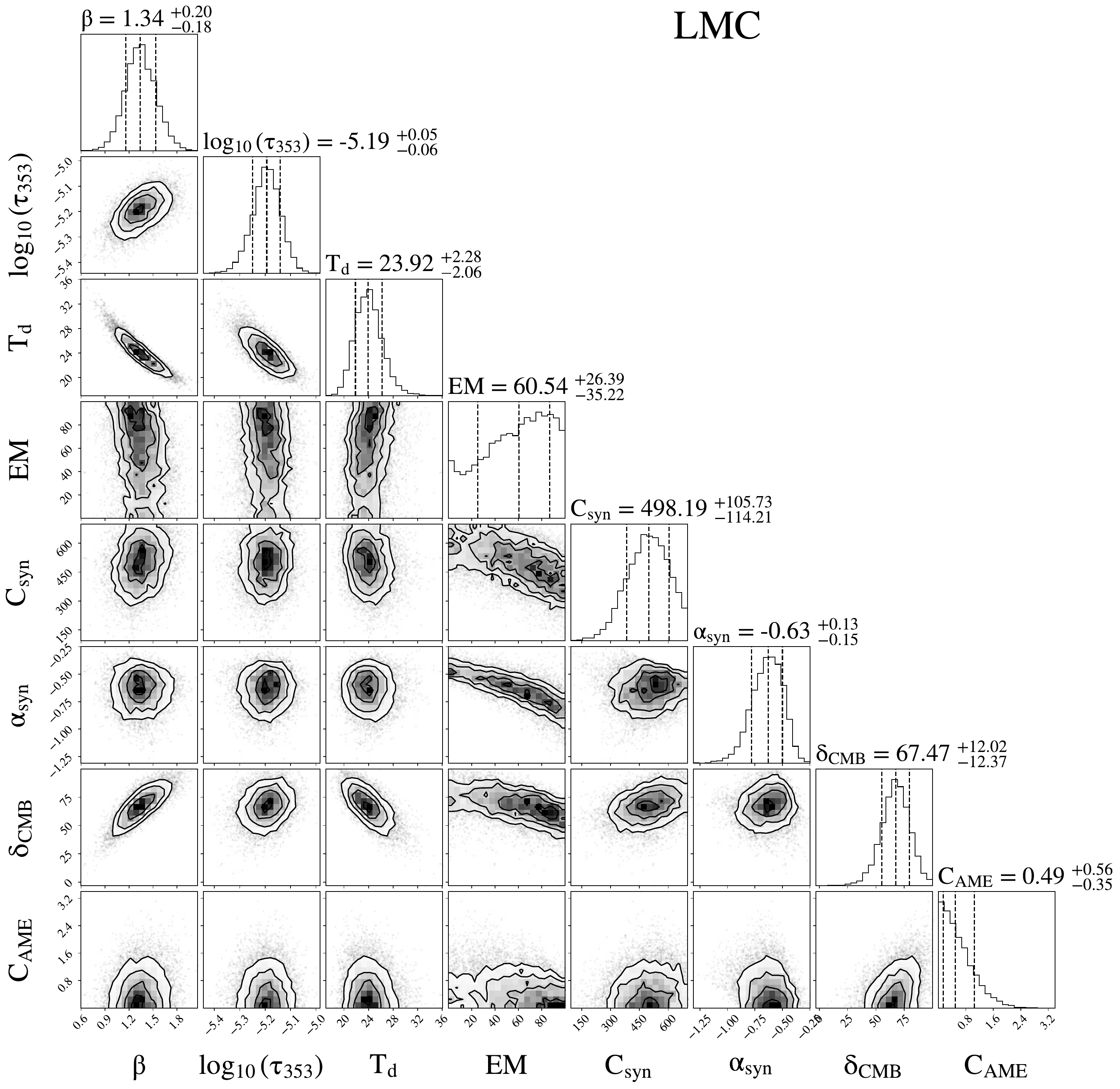}
           \caption{SED of LMC as observed with gray points and without CMB with black points (subtracted from the best model), and radio data in light blue (from Tables \ref{table_radio_data_lmc}) with 30$\%$ increased uncertainties. Data points from the literature \citep{israel_2010,planck_2011} are overlaid in green. The best fit model spectra are overlaid for the global model and individual emission components. On the right, the corner plot displays probability distributions of each model parameter, with the MCMC best-fit values indicated above each histogram.}
    \label{sed_lmc_radio}
\end{figure*}

\end{appendix}

%
%






   
  



\end{document}